\documentclass[11pt]{article}

\usepackage{amsmath,amssymb,tikz}
\usepackage{placeins}
\usetikzlibrary{arrows.meta,decorations.pathmorphing}
\usepackage[margin=1in]{geometry}
\usepackage[colorlinks=true,linkcolor=blue,citecolor=blue,urlcolor=blue]{hyperref}
\let\originalleft\left
\let\originalright\right
\renewcommand{\left}{\mathopen{}\mathclose\bgroup\originalleft}
\renewcommand{\right}{\aftergroup\egroup\originalright}

\numberwithin{equation}{section}

\begin{document}

\hypersetup{pageanchor=false}
\begin{titlepage}
\centering
\vspace*{0.2\textheight}
{\LARGE\bf Single-Minus Graviton Amplitudes\\from Matrix Theory\par}
\vspace{3em}
{\large Alfredo Guevara\(^{1}\),
Alexandru Lupsasca\(^{2,3}\),
Juan Maldacena\(^{1}\),
and Andrew Strominger\(^{2,4}\)\par}
\vspace{3em}
$^1$Institute for Advanced Study, Princeton, NJ, USA \\
$^2$OpenAI, San Francisco, CA, USA \\
$^3$Vanderbilt University, Nashville, TN, USA \\
$^4$Harvard University, Cambridge, MA, USA \\
\vspace{3em}

\begin{abstract}
The $n$-point single-minus graviton amplitudes are nonzero in a special kinematic region that preserves some supersymmetry when embedded in supergravity.
We compute these amplitudes within the BFSS matrix theory by relating them to an index problem considered by A.~Sen.
More precisely, the single-minus amplitudes are obtained from an index that counts BPS states in the Coulomb branch of four-dimensional ${\cal N}=4$ super Yang--Mills.
We explicitly check this relation in a specific kinematic region by computing the index using BPS wall-crossing formulae and matching it to the gravity result.
We further employ BPS wall-crossing to show that the index satisfies a tower of $w_{1+\infty}$ soft theorems in any kinematic regime, again matching gravity.
Finally, we show that for certain kinematics, the single-minus amplitudes can be realized by scattering a graviton off a plane wave.
\end{abstract}

\hypersetup{pageanchor=true}
\end{titlepage}

\tableofcontents
\vspace{3em}

\section{Introduction}
\label{sec:introduction}

The BFSS conjecture identifies M-theory amplitudes with the large-$N$ limit of scattering amplitudes in a supersymmetric matrix quantum mechanics~\cite{BFSS}.
A direct test of the conjecture involves computing a graviton scattering amplitude in the BFSS matrix model, taking the appropriate low-energy limit, and comparing it with the known answer in eleven-dimensional gravity.
In general, this is very difficult, but supersymmetry renders certain special amplitudes tractable.
A notable example is the three-graviton amplitude, which was derived from the matrix model in~\cite{HM3} by relating it to a protected supersymmetric index.

It is natural to wonder whether there are other supersymmetric amplitudes to which a similar trick can be applied.
By a supersymmetric amplitude, we mean an amplitude with a single spinor preserved by all the external momenta, that is, a spinor obeying
\begin{align}
    \Gamma^\mu p^i_\mu \epsilon = 0,
    \quad\text{for all }i.
\end{align}
This condition implies that $p_i\cdot p_j=0$, which is automatic for three-point amplitudes but imposes a nontrivial kinematic restriction for higher-point amplitudes.
  
A natural amplitude of this kind is the four-dimensional single-minus graviton amplitude discussed in~\cite{GLSSW}; see also~\cite{GLSSW-gluon}.
Recall that a single-minus graviton amplitude is zero for generic momenta.
However, \cite{GLSSW} found that in $(2,2)$ signature, the amplitude is nonzero in a special kinematic region where $p_i\cdot p_j=0$ for all particles.
In supergravity, the external particles in this special kinematic configuration preserve a common set of supersymmetries, as in the three-point amplitude.
As a result, the only nonzero amplitudes in this configuration are supersymmetric.

In this paper, we explain how this amplitude can be computed in matrix theory by relating it to a supersymmetric index of the type discussed in~\cite{SenNetwork,Sen}, which involves BPS states for a quarter-supersymmetric $(p,q)$-string web configuration.

We explicitly compare the two results in some special kinematic regions where the index computation has previously been performed, and we find agreement.

Let us give some more details.
In standard spinor-helicity variables [see \eqref{eq:ReviewSpinors} and section~\ref{sec:SMA} for conventions], the single-minus graviton amplitude, computed in~\cite{GLSSW} in ordinary gravity, and extended to ${\cal N}=8$ supergravity in~\cite{BPTW}, is
\begin{align}
    \label{eq:IntroFullAmplitude}
    \mathcal{A}_n(1,2,\ldots,n) 
    =
     2\kappa_{11}^{n-2}(2\pi)^2
     \mathcal{M}_n^{\rm grav} \prod_{a=1}^{n-1}\left[2\pi\delta(\langle an\rangle)\right]
    \delta^2\!\left(\sum_{i=1}^n\tilde{\lambda}^i_{\dot\beta}\right)
    \delta^8\!\left(\sum_{i=1}^n\tilde{\eta}^i_I\right),
\end{align} 
where we choose a little-group gauge in which the first component of every $\lambda^i$ is one,
\begin{align}
    \label{eq:frcho}
    \lambda^i_{\alpha}=\begin{pmatrix} 1 \\ z_i \end{pmatrix},\qquad
    z_i:=\frac{\lambda^i_2}{\lambda^i_1},\qquad
    \langle i j\rangle = z_i-z_j .
\end{align}
The delta functions set all the $z_i$ equal, while the $\tilde{\lambda}^i$ are generic.
Note that such half-collinear kinematics are possible in $(2,2)$ signature but not in $(1,3)$ signature.

We also introduced the fermionic spinor-helicity variables appropriate to ${\cal N}=8$ supergravity.
This kinematic configuration preserves one quarter of the supersymmetries, given by $\lambda^{n\alpha} Q_{\alpha I}$ with $I=1,\ldots,8$. 
The prefactor $\mathcal{M}_{n}^{\rm grav}$ is a function only of the brackets $[ij]=\epsilon^{\dot\alpha\dot\beta}\tilde{\lambda}^i_{\dot\alpha}\tilde{\lambda}^j_{\dot\beta}$.
It is characterized by kinematic chambers, separated by singularities across which the first derivative of the amplitude may be discontinuous.
Within each chamber, $\mathcal{M}_{n}^{\rm grav}$ is simply a polynomial of degree $n-2$ in the brackets $[ij]$.
A simple chamber that we will consider in this paper is defined by 
\begin{align}
    \label{eq:IntroDecayChamber}
    \mathcal{R}_{n,n-1}:\qquad
    \tilde{z}_1<\cdots<\tilde{z}_{n-2}<\tilde{z}_n<\tilde{z}_{n-1},\qquad
    {\rm where}\qquad
    \tilde{z}_i := \frac{\tilde{\lambda}^i_{\dot 2}}{\tilde{\lambda}^i_{\dot 1}},\qquad
    \tilde{\lambda}^i_{\dot\beta}=
    \begin{pmatrix}
        \tilde{\lambda}^i_{\dot 1} \\[2pt]
        \tilde{\lambda}^i_{\dot 2}
    \end{pmatrix}.
\end{align}
We will also take the first $n-1$ particles to have positive energy and to be incoming, and the last particle to have negative energy and to be outgoing.
In this case, \cite{GLSSW} found that $\mathcal{M}_n^{\rm grav}$ collapses within this chamber to the $(n-2)$-fold product
\begin{align}
    \label{eq:IntroProduct}
    \mathcal{M}^{\rm grav}_{n}\big|_{\mathcal{R}_{n,n-1}}=\prod_{a=1}^{n-2}S_a,\qquad
    S_a:=\frac{1}{2}\sum_{j\ne a}|[aj]|
    =\sum_{b=a+1}^{n-1}[ba].
\end{align}
For example,
\begin{align} \label{Intro34pt}
    \mathcal{M}_3^{\rm grav}\big|_{\mathcal{R}_{3,2}}=[21],\qquad
    \mathcal{M}_4^{\rm grav}\big|_{\mathcal{R}_{4,3}}=([21]+[31])[32].
\end{align}

By adding seven extra dimensions and considering particles with zero external momentum along them, we can also view \eqref{eq:IntroFullAmplitude} as an amplitude in eleven dimensions, written in terms of four-dimensional variables.
Eleven-dimensional amplitudes that preserve a quarter of the supersymmetry have external momenta lying in a four-dimensional subspace.

In this paper, we precisely reproduce this product from BFSS matrix theory (reviewed in section~\ref{sec:ScatteringBFSS}), viewing the corresponding scattering process as an eleven-dimensional amplitude with special four-dimensional kinematics.
The idea will be the same as in~\cite{HM3}.
Namely, we compactify an extra dimension, so that we now have two compact dimensions: the lightlike direction $x^-$ and one of the spatial dimensions, $x^9$.
Then we find that
\begin{align}
    \tilde{\lambda}^i_{\dot\alpha}=
    \begin{pmatrix}
        N_i/R_-\\[2pt]
        \sqrt{2}n_i/R_9
    \end{pmatrix},
\end{align}
and the computation reduces to one in matrix string theory~\cite{TaylorCompact,BanksSeiberg,MatrixString}.
  
\begin{figure}[htbp]
    \begin{center}
    \includegraphics[width=.6\textwidth]{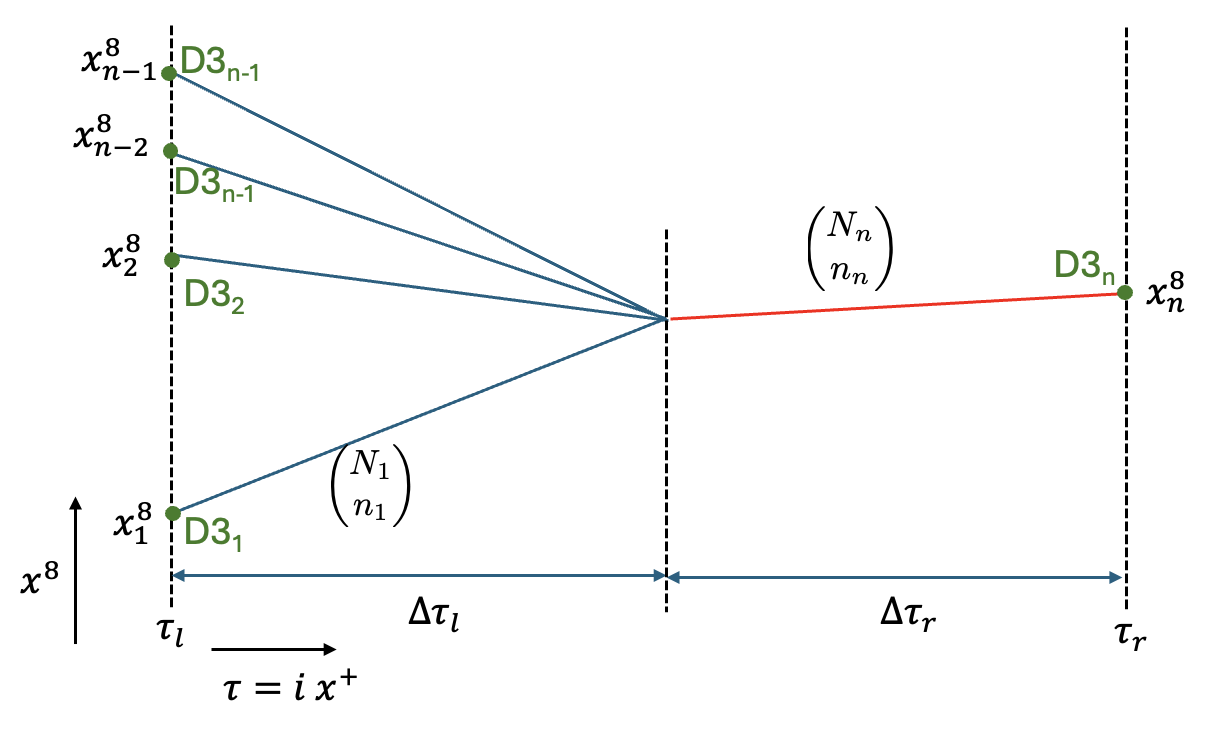}
    \end{center}
    \caption{Diagram of the scattering configuration, shown in the kinematic regime \eqref{eq:IntroDecayChamber}.
    Given the particle momenta $(N_i,n_i)$, supersymmetry determines the slopes of the lines shown in the $(ix^+,x^8)$ plane.
    We arrange the positions of the D3-branes so that all the lines meet at a single point.}
    \label{fig:Scattering}
\end{figure}

We then relate the scattering configuration to a string network suspended between D3-branes; see Fig.~\ref{fig:Scattering}.
Each external line corresponds to a string with $N_i$ units of D1-brane charge and $n_i$ units of F1-string charge.
This can also be viewed as a configuration for a $\mathsf{U}(N)$ gauge theory on an interval with suitable boundary conditions at its ends.
We can therefore view the configuration as both an ``open-string'' channel and a ``closed-string'' channel.
The computation in the open-string channel is an index because we can view the direction $\tilde{x}^9$ (which is the T-dual of $x^9$) as Euclidean time with an insertion of $(-1)^F$, since the fermions have periodic boundary conditions.
The computation in the closed-string channel can be expressed in terms of the amplitude, allowing us to read off the expression for $\mathcal{M}_n^{\rm BFSS}$, which is essentially the index.
More precisely, we will argue in section~\ref{sec:Connection} that 
\begin{align}
    \label{eq:IndexM}
    \boxed{\mu^{2-n}\mathcal{M}_n^{\rm BFSS}=\mathcal{I}_n
    =-\frac{1}{[2(n-1)]!}\,{\rm Tr}\left[(-1)^F \frac{(2J)^2}{2} (2I_3)^{2(n-1)}\right]}\,.
\end{align}
Here, $\mu=\frac{\sqrt{2}}{R_-R_9}$, $J$ is the spin in the D3-brane directions, and $I_3$ is the generator of the $\mathsf{SU}(2)_+$ rotation along four of the directions orthogonal to the D3-brane (these insertions are necessary to absorb various fermion zero modes).
One might have expected additional momentum-independent factors to enter this equation, but with our normalizations, these factors exactly cancel out, leaving precisely \eqref{eq:IndexM} as written above.
The argument for \eqref{eq:IndexM} does not rely on the choice of chamber. 

Here, the index is a function of the brackets
\begin{align}
    \label{eq:GammaDefinition}
     \langle\Gamma_i,\Gamma_j\rangle = N_i n_j - N_j n_i,
     \qquad{\rm with}\qquad
     \Gamma_i=\begin{pmatrix} N_i \\ n_i \end{pmatrix}
     \qquad{\rm and}\qquad
     [ij] =-\mu\langle\Gamma_i,\Gamma_j\rangle.
\end{align}
Each $\Gamma_i$ is the charge of the BPS state under the gauge group $\mathsf{U}(1)_i$ on the $i^{\rm th}$ D3-brane; see Fig.~\ref{fig:Scattering}.
Using the method explained in~\cite{Sen}, we compute the index in section~\ref{sec:IndexComputation} and find precise agreement with \eqref{eq:IntroProduct} in the particular chamber \eqref{eq:IntroDecayChamber}.
It is likely that the correspondence \eqref{eq:IndexM} holds in more general chambers and involves delicate cancellations, but we leave such checks for future work.
Some general comments on this correspondence and further directions are discussed in section~\ref{sec:Comparison}.

Gravitational scattering amplitudes are constrained by an ${\cal L}w_{1+\infty}$ tower of soft theorems~\cite{Guevara:2021abz,Strominger:2021mtt}.
For single-minus amplitudes, an $n$-point amplitude at any one point in any chamber determines all the higher-point amplitudes within the same chamber via the all-orders soft theorem~\cite{GLSSW}
\begin{equation} \label{eq:wsoft-ward-intro}
 {\cal M}_n(1,\ldots,n-1,s)
 =\frac{1}{2}
 \sum_{j=1}^{n-1}
  {\lvert}[sj] {\rvert}\,
 {\cal M}_{n-1}(1,\ldots,j+s,\ldots,n-1).
\end{equation}
This identity is to be understood term-by-term in an expansion in $|s]$; generically, for sufficiently large $|s]$, a chamber wall is encountered and the identity breaks down.\footnote{MHV amplitudes are known to follow everywhere from ${\cal L}w_{1+\infty}$~\cite{Guevara:2025tsm}.
This may also be true for the single-minus amplitudes, but the chamber wall-crossing conditions have not been analyzed in this light.}
In section~\ref{sec:Soft}, we rederive \eqref{eq:wsoft-ward-intro} within BFSS as a BPS wall-crossing identity for the index (not to be confused with chamber wall-crossing!).
In the decay chamber, this recursion relation reproduces all the single-minus amplitudes from the three-point seed and hence confirms the numerical match found in section~\ref{sec:IndexComputation}.
Since \eqref{eq:wsoft-ward-intro} is valid in any kinematic chamber, it supports the idea that the index and the amplitude agree for all kinematics.

Finally, in section~\ref{sec:pp-wave-check}, we consider a further kinematic restriction in which we set all but two of the $N_i$ to zero.
In this case, the matrix-model computation reduces directly to the gravity computation.
In fact, the gravity computation in this regime is of intrinsic interest because it can be viewed as the scattering of a graviton with nonzero $p_-$ off a complex plane wave created by the $n-2$ gravitons with zero $p_-$.

\section{Single-minus graviton amplitudes in supergravity}
\label{sec:SMA}

In this section, we describe our conventions, write the amplitude more explicitly, and explain its supersymmetry.
Readers uninterested in the details can proceed directly to the final formula, \eqref{AGravFin}.

\subsection{Conventions and supersymmetry}
 
In this paper, we consider an eleven-dimensional kinematic configuration of momenta that preserves one quarter of the 32 supersymmetries of M-theory.
Such a configuration can be Lorentz-transformed to one that lies in a single four-dimensional subspace.\footnote{There are supersymmetric configurations of momenta that cannot be Lorentz-transformed to a four-dimensional subspace.
However, such configurations preserve strictly less than one quarter of the supersymmetry.}
For that reason, we use four-dimensional spinor-helicity notation to describe the external particles and their polarizations.
We follow conventions similar to those in~\cite{HM3}.

We choose eleven-dimensional coordinates $(x^+,x^-,x^9,x^8,x^7,\ldots,x^1)$.
The four-dimensional subspace is parameterized by the coordinates $(x^+,x^-,x^9,x^8)$, with metric 
\begin{align}
    ds^2=-2\,\mathrm{d}x^+\,\mathrm{d}x^-+2\,\mathrm{d}z\,\mathrm{d}\bar{z},\qquad
    z:=\frac{x^9+ix^8}{\sqrt{2}}.
\end{align}
  
Graviton momenta are chosen in a frame such that $p_+=p_{\bar{z}}=0$.
In these coordinates,
\begin{align} \label{MomVal}
    p^i_{z}=\frac{p^i_9 -i p^i_8}{\sqrt{2}}
    =\sqrt{2}\,p_i^9,\qquad
    p^i_{\bar{z}}=\frac{p^i_9 +i p^i_8}{\sqrt{2}}
    =0.
\end{align} 
The associated spinors are\footnote{This differs slightly from~\cite{HM3}, where all $\tilde{\lambda}^i$ were chosen to be equal.}
\begin{align}
    \label{eq:ReviewSpinors}
    \lambda^i=\begin{pmatrix} 1 \\ 0 \end{pmatrix},\qquad
    \tilde{\lambda}_i=
    \begin{pmatrix}
        -p^i_{-} \\[2pt]
        \sqrt{2}\,p^i_9
    \end{pmatrix},\qquad
    p^i_{\alpha\dot\alpha}=\lambda^i_{\alpha}\tilde{\lambda}^i_{\dot\alpha}
    =
    \begin{pmatrix}
        -p_- & p_z \\
        -p_{\bar{z}} & p_+
    \end{pmatrix}.
\end{align}
The momentum matrix obeys
$p^2=2\det p$. In ${\cal N}=8$ supergravity notation, we have the supercharges $Q_{\alpha I}$ and $\tilde{Q}_{\dot\beta}^{J}$ with $I,J=1,\ldots,8$.
Acting on a massless on-shell particle, they obey the algebra
\begin{align}
    \{ Q_{\alpha I},\tilde{Q}_{\dot\beta}^{J} \} = \lambda_{\alpha}\tilde{\lambda}_{\dot\beta}\delta_{I}^J.
\end{align} 
For each particle, choose $w_\alpha$ and $\tilde{w}_{\dot\beta}$ such that $\langle w,\lambda\rangle=1$ and $[\tilde{w},\tilde{\lambda}]=1$.
We can then define the multiplet 
\begin{align}
    \label{eq:GeneralMultiplet}
    |\tilde{\eta} \rangle = e^{b^{\dagger I} \tilde{\eta}_{I} }|+\rangle,\qquad
    b_I |+\rangle = 0,\qquad
    {\rm with }\qquad
    b^{\dagger I}\equiv {\tilde{w}}^{\dot \beta} \tilde{Q}_{\dot \beta }^{I },\qquad
    b_I \equiv w^{\alpha } Q_{\alpha I },
\end{align} 
where $|+\rangle$ is the positive-helicity graviton.
With these definitions, we find that the supercharges acting on the amplitude have the form 
\begin{align}
    \label{eq:SupE}
    Q_{\alpha I } = \sum_i \lambda^i_{\alpha } \tilde{\eta}^i_{I},\qquad
    \tilde{Q}_{\dot \beta }^{ I } = \sum_i {\tilde{\lambda}}^i_{\dot \beta } \frac{\partial}{\partial \tilde{\eta}^i_{I}}.
\end{align} 
We see that if all the $\lambda^i \propto \lambda$ are parallel, then the amplitude will automatically preserve a quarter of the supercharges, the ones given by 
\begin{align}
    \label{eq:PreservedSUSY}
    \lambda^{\alpha}Q_{\alpha I }.
\end{align} 
Notice from \eqref{eq:SupE} that these supercharges are identically zero. 
Furthermore, we can also see that a fully supersymmetric amplitude is given by\footnote{We will use
$\kappa_{11}^{2}=8\pi G_{11}=(2\pi)^8 \ell_p^9/2$  and
$g_{MN}=\eta_{MN}+2\kappa_{11}h_{MN}$. With respect to~\cite{GLSSW} we strip the amplitude phase $i^{n-2}$ and normalize the delta functions as $\int dx\delta (x)=1$ (as opposed to $2\pi$ in that reference).}
\begin{align}
    \label{eq:SUSA}
    {\cal A }_n = 2\kappa_{11}^{n-2} (2\pi)^9\mathcal{M}_n^{\rm grav} \left[ \prod_{i=1}^{n-1} 2\pi \delta\!( \langle i, n \rangle ) \right]\delta^8\!\left( \sum_{i=1}^n \tilde{\eta}_{I }^i \right)  \delta^2\!  \left( \sum_{i=1}^n \tilde{\lambda}^{i}_{\dot \beta } \right) \delta^7\!\left(\sum_{i=1}^n \vec k_i\right),
    \quad{\rm with}\quad
    \lambda^i_\alpha=\begin{pmatrix} 1 \cr z_i \end{pmatrix},
\end{align} 
where we have used the freedom to rescale each $\lambda_i$.
The delta functions set all $z_i$ to be equal, and by performing a Lorentz transformation we can then set them all to zero.
The last delta function takes into account the momenta $\vec k$ along the other seven dimensions.
Here $\mathcal{M}^{\rm grav}$ is a Lorentz-invariant function of the $\tilde{\lambda}_{\dot\beta}^i$. Appendix~\ref{app:three-point-normalization} provides a dictionary between this convention and the textbook superamplitude for $n=3$.

It is easy to see that \eqref{eq:SUSA} preserves all the supersymmetries \eqref{eq:SupE}.
We can also check that this is the only expression with precisely eight $\tilde{\eta}_I$ that preserves all the supersymmetries in \eqref{eq:PreservedSUSY}.
Namely, for generic $\lambda^i$ we need at least 16 factors of $\tilde{\eta}_I$, which corresponds to the usual MHV amplitude with two negative-helicity polarizations.

Note that \eqref{eq:SUSA} cannot have any higher order corrections in $G_N$, since such corrections would be accompanied by ordinary Mandelstam invariants which all vanish for this amplitude.  In other words, functions of only $p_-$ and $p_z $ in \eqref{MomVal} and homogeneity degree two cannot be Lorentz invariant. 

In order to make contact with the matrix theory discussion, we compactify both $x^-$ and $x^9$, leading to quantized momenta
\begin{align}
    \label{eq:ReviewMomenta}
    -p^i_{-}=\frac{N_i}{R_-},\qquad
    p^i_9=\frac{n_i}{R_9},\qquad
    x^-\sim x^-+2\pi R_-,\qquad
    x^9\sim x^9+2\pi R_9 .
\end{align}
It is useful to define 
\begin{align}
    \Gamma_i=(N_i,n_i),\qquad
    \langle\Gamma_i,\Gamma_j\rangle=N_i n_j-N_j n_i,
\end{align}
which will play the role of an electromagnetic charge-lattice vector later in the paper; for now, this is just a definition.
We will take the first $n-1$ particles to be incoming and the $n^{\rm th}$ one to be outgoing.
Thus, we define 
\begin{align}
    \label{eq:NVal}
    N \equiv \sum_{i=1}^{n-1} N_i,\qquad
    N_n \equiv - N,\qquad
    n_n = - \sum_{i=1}^{n-1} {n_i},
\end{align} 
where $N_i>0$ for $ 1 \leq i \leq n-1$.
Note that  
\begin{align}
    \label{eq:ReviewDictionary}
    \tilde{\lambda}_i
    =\begin{pmatrix}
        N_i/R_-\\[2pt]
        \sqrt{2} n_i/R_9
    \end{pmatrix},\qquad
    [ij]=-\mu\langle\Gamma_i,\Gamma_j\rangle,
\end{align}
with $\mu=\sqrt2/(R_-R_9)$.
Hence, a dimensionless lattice multiplicity becomes a spinor bracket after multiplication by one power of $\mu$.
With the orientation fixed above, the displayed spinors give $[ij] = -\sqrt{2} (N_i n_j - N_j n_i) / (R_- R_9)$, so both the sign and the factor $\mu$ in \eqref{eq:ReviewDictionary} follow directly.

Notice that in writing \eqref{eq:SUSA}, we have only used supersymmetry.
Therefore, after the compactification \eqref{eq:ReviewMomenta}, or in the matrix model, supersymmetry and the manifest symmetries of the matrix model imply the expression in \eqref{eq:SUSA}, except for the substitution $\mathcal{M}_n^{\rm grav}\to\mathcal{M}_n^{\rm BFSS}$, where $\mathcal{M}_n^{\rm BFSS}$ is a function of the $N_i$ and $n_i$ that, in principle, need not be Lorentz-invariant.
Of course, a consistency check will be that we obtain a Lorentz-invariant $\mathcal{M}^{\rm BFSS}$, as expected from the matrix-theory soft theorems~\cite{HMsoft}.
In particular, note that supersymmetry implies the first and second sets of delta functions in \eqref{eq:SUSA}.

\subsection{Rewriting the superamplitude in terms of a new ``vacuum''}
\label{sec:M2Vacuum}

In preparation for the matrix-model discussion, it is useful to express the amplitude in terms of a different ``vacuum'' instead of the state $|+\rangle$ in \eqref{eq:GeneralMultiplet}.
This is because, in order to make the connection to the index, it will be convenient to view the amplitude as arising from a correlation function of sources.
These sources will be Euclidean M2-branes in M-theory that are extended along the directions 123.
Such branes preserve the supersymmetries given by
\begin{align}
    \label{eq:SusyM2}
    i \Gamma^{123} \epsilon = \epsilon.
\end{align} 
These sources can be viewed as producing certain modes of the massless fields that are simply the long-range fields produced by the M2-branes.
Such modes are annihilated by \eqref{eq:SusyM2}.
If we produce a field mode with some momentum, then we want to choose a new ground state, called $|m\rangle$, that is annihilated by these supersymmetries.
The restriction \eqref{eq:SusyM2} also fixes the product of chiralities in the four dimensions spanned by $x^+,x^-,x^9,x^8$ (which distinguishes the $Q$ and $\tilde{Q}$ supercharges) and the chirality in the four dimensions spanned by $x^7,x^6,x^5,x^4$ which will split the eight $I$ indices into four $K$ and four $J$ indices distinguished by this chirality. 
In other words, we separate
\begin{alignat}{3}
    \label{SUPDec}
    Q_{\alpha I } & \to Q_{\alpha K },~ Q_{\alpha J } \qquad & Q_{\alpha K} &:~ (i\Gamma^{123} , \Gamma^{4567}) = ( -, +) \qquad Q_{\alpha J} & :~ (+ , - ) \\
    \tilde{Q}_{\dot \beta }^{I} & \to \tilde{Q}_{\dot \beta}^{K },~~ \tilde{Q}_{\dot \beta}^{J} \qquad  & \tilde{Q}_{\dot \beta}^{K } &:~ (i\Gamma^{123} , \Gamma^{4567}) = ( +, +) \qquad \tilde{Q}_{\dot \beta}^{J} & :~ (-, - )
\end{alignat}
where we decompose the eight indices $I$ into two sets of four denoted by $K$ and $J$. 
Having thus separated the supercharges, we can define the multiplet using the new vacuum 
\begin{align}
    \label{eq:MVacDef}
    |\eta^K , \tilde{\eta}_J \rangle = \int d^4 \tilde{\eta}^K e^{\tilde{\eta}^K \eta_K} |\tilde{\eta}_I\rangle
    = e^{\eta^K b_K} e^{\tilde{\eta}_J b^{\dagger J}} |m \rangle,\qquad
    |m\rangle = \prod_{K=1}^4 b^{\dagger K } |+ \rangle.
\end{align} 
Note that $|m\rangle$ is annihilated by the supercharges $Q_{\alpha J}$ and $\tilde{Q}^K_{\dot\beta}$ that are preserved by the M2-branes.

By a simple Grassmann Fourier transformation, it is straightforward to write the amplitude in terms of this new way of describing the multiplet: 
\begin{align}
    {\cal A }_n = 2\kappa_{11}^{n-2} (2\pi)^9 \mathcal{M}_n^{\rm grav} \left[ \prod_{i=1}^{n-1} 2\pi \delta( \langle i ,n \rangle ) \right] \delta^2\!\left( \sum_{i=1}^n \tilde{\lambda}^{i}_{\dot \beta } \right) \delta^7\!\left(\sum_{i=1}^n \vec k_i\right) \delta^4\!\left( \sum_{i=1}^n \tilde{\eta}_{J }^i \right) \prod_{K=1}^4 \left( \sum_i \partial_{\eta^i_K} \right) \prod_{i=1}^n \delta^4\!\left(\eta^i_K\right).
\end{align} 
The amplitude is totally permutation symmetric.
The single-minus amplitude, with the minus in the $n^{\rm th}$ leg, is obtained from the following component of the Grassmann expansion, on which we will focus from now on:
\begin{align}
    \label{eq:AMpInt}
    {\cal A }_n = 2\kappa_{11}^{n-2} (2\pi)^9  {\cal M }_n^{\rm grav} \left[ \prod_{i=1}^{n-1} 2\pi \delta( \langle i ,n \rangle )  \delta^4 \left( \eta^K_i \right)\right] \delta^2 \left( \sum_{i=1}^n \tilde{\lambda}^{i}_{\dot \beta } \right)  \delta^4 \left(  \tilde{\eta}_{J }^n \right) \delta^7 ( \sum_i \vec k_i)   ~,~~~ {\rm with } ~~\lambda^i_\alpha = \begin{pmatrix} 1 \cr z_i \end{pmatrix}~.
\end{align} 
The last equality in \eqref{eq:AMpInt} sets the scale of all $\lambda^i$ equal.
As a final rewriting, let us note that in the Lorentz frame where $z_n=0$ we can rewrite
\begin{align}
    \prod_{i=1}^{n-1} \delta( \langle i ,n \rangle ) = \prod_{i=1}^{n-1} \delta( z_i )
    = \prod_{i=1}^{n-1} \frac{\sqrt{2}}{R_-} N_i \delta( -ip_8^i - p_9^i),\qquad
    p_9^i = \frac{n_i}{R_9},\qquad
    \lambda^n_{\alpha} = \begin{pmatrix}1\\[2pt]0\end{pmatrix},
\end{align} 
where the delta functions for the $p_8^i$ force them to take their supersymmetric value.\footnote{A delta function of a complex quantity is not well-defined; here, $-ip_8^i$ are the real oriented integration variables.}

We could also write the amplitude in the canonical normalization, as detailed in~\cite{HM3},
\begin{align}
    |p_i\rangle_c=\frac{1}{2\pi\sqrt{2|N_i|R_9}}|p_i\rangle,
\end{align}
resulting in
\begin{align}\label{AGravFin}
{\cal A}^c_n
&{=}
{\frac{R_-\sqrt{R_9}}{(2\pi)^{n-9}}}
\frac{\kappa_{11}^{\,n-2}}{\sqrt{N}} {\cal M }^{\rm grav}
\left[ \prod_{i=1}^{n-1}
{\frac{2\pi }{R_-\sqrt{R_9}}}\sqrt{N_i}
{\delta\!\left(-ip_8^i-\frac{n_i}{R_9}\right)}
\delta^4 \left( \eta^K_i \right)\right]
\\[-2pt]
&\quad\times
{\delta_{\sum_iN_i,0}\delta_{\sum_in_i,0}}
\delta^4 \left( \widetilde\eta_J^n \right)
\delta^7 \left( \sum_i \vec k_i\right),
\qquad \lambda^i_\alpha=\begin{pmatrix}1\\[2pt]0\end{pmatrix}.
\end{align}
Notice that for $n=3$, the three-point amplitude, the total number of delta functions for momentum conservation is indeed eleven, as expected. Note, in particular, that overall momentum conservation in the eighth direction follows from these relations. For $n>3$, we have additional delta functions that restrict the amplitude to a subspace. The choice of $\lambda_n$ in \eqref{AGravFin} ensures that $p_8^n=i p_9^n$ and is compatible with the compactification of $x^-$ and $x^9$.

We should point out that all we have done so far is rewrite the standard graviton amplitude \eqref{eq:SUSA} for the special case of quantized momenta $p_-$ and $p_9$, and in terms of fermionic variables well adapted to what we will later do in the matrix model.

\section{Scattering problem in matrix string theory}
\label{sec:ScatteringBFSS}

BFSS is the maximally supersymmetric $\mathsf{U}(N)$ quantum mechanics of nine
Hermitian matrices $X^I(x^+)$, their fermionic partners, and the
one-dimensional gauge field $A_+$~\cite{BFSS}. It is proposed to describe
the DLCQ sector of M-theory carrying $N$ units of longitudinal
momentum~\cite{SusskindDLCQ,SeibergDLCQ}. In our conventions, this is
\begin{align}
  -p_-=\frac{N}{R_-}.
  \label{eq:review-bfss-longitudinal-momentum}
\end{align}
The diagonal entries of the $X^I$ are the transverse positions of the
D0-branes. When diagonal entries
approach each other, the off-diagonal entries become light and generate interactions. The $\mathsf{SU}(N)$ part describes the
relative motion and the decoupled $\mathsf{U}(1)$ describes the center of mass.
The large-$N$ limit decompactifies the lightlike circle and is the limit in
which the model is conjectured to recover uncompactified M-theory~\cite{BFSS}.

When one transverse direction $x^9$ is also compactified, the appropriate description is obtained by T-dualizing
\cite{TaylorCompact,MatrixString}.
Equivalently, the BFSS quantum mechanics becomes maximally supersymmetric $1+1$-dimensional $\mathsf{U}(N)$ Yang--Mills theory:
\begin{align}
  \left.
  \text{BFSS on }S^1_9(R_9)
  \right.
  \quad\overset{T_9}{\longleftrightarrow}\quad
  \left.
  \mathsf{U}(N)\text{ SYM}_{1+1}\text{ on }\tilde{S}^1_9(\tilde{R}_9)
  \right.,
  \qquad
  \tilde{R}_9=\frac{\ell_p^3}{R_-R_9}.
  \label{eq:review-bfss-matrix-string-duality}
\end{align}
Here $\ell_p$ is the eleven-dimensional Planck length.  Under this
duality, the integer momentum $n_e$ along $x^9$ becomes electric flux in
the two-dimensional gauge theory; see~\cite{MatrixString}.

We will consider the scattering configuration in \eqref{eq:NVal}. In the matrix-string, or BFSS-like, description, the incoming state corresponds to $n-1$ blocks of sizes $N_i$ and with electric fluxes $n_i$, coming together and becoming a block of size $N=\sum_{i=1}^{n-1}N_i$. For coprime $(N_i,n_i)$, the corresponding $\mathsf{SU}(N_i)$ sector has an
isolated supersymmetric ground state~\cite{WittenBoundStates}. These are our asymptotic states; more precisely, they form an asymptotic multiplet built from the fermion zero modes living in $\mathsf{U}(1)_i$ for each $\mathsf{U}(N_i)\sim \mathsf{U}(1)_i\times \mathsf{SU}(N_i)$.

Using supersymmetry, the fact that we focus on an amplitude preserving a quarter of the supersymmetry, and the residual part of the Lorentz group compatible with the compactification of $x^-$ and $x^9$ (which is a manifest symmetry of the matrix string), we can transform the problem to a region where only $p_8$ is nonzero
and imaginary, while all other continuous momenta of the external matrix-string states are set to zero.

This means that we can write the matrix-string scattering amplitude as in \eqref{AGravFin}:
\begin{align} \label{AmplBFSS}
{\cal A}^{c \rm BFSS}_n
&{=}
{\frac{R_-\sqrt{R_9}}{(2\pi)^{n-9}}}
\frac{\kappa_{11}^{\,n-2}}{\sqrt{N}} {\cal M }_n^{\rm BFSS}
\left[ \prod_{i=1}^{n-1}
{\frac{2\pi}{R_-\sqrt{R_9}}}\sqrt{N_i}
{\delta\!\left(-ip_8^i-\frac{n_i}{R_9}\right)}
\delta^4 \left( \eta^K_i \right)\right]
\nonumber \\[-2pt]
&\quad\times
{\delta_{\sum_iN_i,0}\delta_{\sum_in_i,0}}
\delta^4 \left( \widetilde\eta_J^n \right)
\delta^7 \left( \sum_i \vec k_i\right),
\qquad \lambda^n_\alpha=\begin{pmatrix}1\\[2pt]0\end{pmatrix}.
\end{align}
except that we have replaced $\mathcal{M}^{\rm grav}_n\to\mathcal{M}^{\rm BFSS}_n$. Now $\mathcal{M}^{\rm BFSS}$ is a function of $N_i$ and $n_i$ that need not be Lorentz invariant. However, our final result will be Lorentz invariant, which is a nontrivial check.  

\section{The connection between the amplitude and the index}
\label{sec:Connection}

In this section, we present an argument relating the above amplitudes to an index computation. This is a simple extension of the argument in~\cite{HM3} to $n$ particles.

The basic idea is the following. The scattering problem that we need to consider for \eqref{AmplBFSS} involves $n-1$ matrix blocks merging into a single block as we propagate in time, or $x^+$. It is convenient to view this as a process in the Euclidean version of the theory, where $x^+=i\tau$ and $\tau$ is real Euclidean time. Notice that $p_8$ was set to an imaginary value in the supersymmetric amplitude. This means that $x_i^8\propto x^+p_8^i/N_i\propto\tau p_9^i/N_i$, which is real. Therefore, in the $(\tau,x^8)$ plane, the trajectories of the external particles are lines with slopes proportional to $n_i/N_i$. The supersymmetric $1+1$-dimensional theory we are considering arises on the worldvolume of D1-branes. The electric field can be viewed as giving rise to fundamental-string charge. Thus, the block of rank $N_i$ with $n_i$ units of electric field can be viewed as a string with $(N_i,n_i)$ units of D1 and F1 charge. Since the strings are slanted in the $(\tau,x^8)$ plane by an amount proportional to their $n_i$ charge, the configuration is supersymmetric. Such BPS string webs were studied by Sen~\cite{SenNetwork,Sen}. Strictly speaking, we are considering the field-theory limit of these networks from the point of view of the worldvolume of the D1-branes. However, this distinction is not important for our discussion of the index below, precisely because it is an index.

It is convenient to have such strings start and end on D3-branes, as in figure \ref{fig:Scattering}. These D3-branes are the matrix-string version of the M2-branes that we mentioned in section~\ref{sec:M2Vacuum}, when we discussed the eleven-dimensional M-theory amplitudes. This implies that the worldvolume theory on the D1-branes, the two-dimensional super Yang--Mills theory, lives on a finite cylinder $S^1\times I$, where $S^1$ is spanned by the T-dual coordinate $\tilde{x}^9$ and $I=[\tau_l,\tau_r]$. We impose supersymmetric Nahm-pole boundary conditions at the endpoints of the interval~\cite{Nahm,Diaconescu}.

The main tool that we will use is the fact that we can view this cylinder computation in the ``open string'' channel, where the circle is Euclidean time, as a trace over the Hilbert space on the interval. We have an insertion of
$(-1)^F$, since the fermions are periodic on the circle. This is a Witten index~\cite{WittenIndex}, because the whole configuration is preserving some supersymmetries. In addition, we can view the same cylinder as a propagation along the ``closed string'' channel where the computation becomes an overlap between two boundary states, which, as we will argue, can be computed in terms of the amplitude. In other words, 
\begin{align}
  {\cal I}_{\rm naive} = \operatorname{Tr}_{{\cal H}_{\rm o}}
  \!\left[
    (-1)^F e^{-2\pi\tilde{R}_9(H_{\rm o}-Z_{\rm o})}
  \right]
  =
  \langle B_l|
  e^{-(\tau_r-\tau_l)(H_{\rm c}-Z_{\rm c})}
  |B_r\rangle ,
  \label{eq:review-open-closed}
\end{align}
where $H_{o,c}$ and $Z_{o,c}$ are open and closed channel Hamiltonians and central charges. The central charges appear because we have some non-trivial electric fields and field gradients and we only want to consider the energy above that supersymmetric state.

Though we sometimes use the language of D3-branes and strings stretched between them, we can phrase the full computation in terms of the $1+1$-dimensional super Yang--Mills theory.
Note that we can phrase these boundary conditions purely in the language of the two-dimensional field theory.  
The boundary conditions on the left side split into boundary conditions for $n-1$ different blocks of size $N_i$. These are separated along the $8^{th}$ direction. In order to make the link to the amplitude, 
 we fix these positions   in such a way that the lines all meet at a point; see figure \ref{fig:Scattering}.   
 We will discuss more general positions later. 
 In addition, each block is also carrying  electric fluxes $n_i$.
At the right end there is a single
  block with $N= \sum_i N_i$ and electric field $n = \sum_i n_i$.

It turns out that the naive index \eqref{eq:review-open-closed} vanishes, due to the presence of fermion zero modes. In order to get a nonzero quantity, we will need to absorb many fermion zero modes. 
The vanishing of \eqref{eq:review-open-closed} is good from the amplitude point of view. It means that no potential is generated when we have separated branes.  Moreover, Sen argued that any index for a connected string network for $n>3$ can be non-zero if the branes all lie on the same two-dimensional transverse plane. For our supersymmetric amplitudes, this is automatic once we take into account that only $p_8$ is nonzero and the rest of the momenta are zero. This implies that the D3 branes are only separated in the $x^8$ and $\tau = i x^+$ directions.  

More technically, it means that if we set all $\eta^K$ and $\tilde{\eta}_J$ variables to zero in \eqref{AmplBFSS}, then the amplitude vanishes. Those factors of $\eta^K$ and $\tilde{\eta}_J$ correspond to insertions of the supercharges acting on the $|m\rangle$ vacuum, and they absorb precisely all the fermion zero modes present in the naive index \eqref{eq:review-open-closed}.

In order to organize the treatment of the fermion zero modes, it is convenient to introduce a new index-like quantity~\cite{Sen}
\begin{align}\label{Indz}
{ B}_2(\nu) = -{\rm Tr}[ (-1)^F \frac{(2J)^2}{2!} e^{ 2 \nu I_3} e^{ -2 \pi \tilde{R}_9   (H_o-Z_o) }]
\end{align} 
where $J$ is the angular momentum in the 123 directions and $I_3$ is a $\mathsf{U}(1) \subset \mathsf{SU}(2)_+ \subset \mathsf{SO}(4) $, where $\mathsf{SO}(4) $ acts in the 4567 directions. Note that a spinor   decomposes as  
\begin{align}\label{SO4}  
(2,1)_+ \oplus (1,2)_- ~,~~~{\rm under} ~~~~ \mathsf{SO}(4)_{4567} \sim \mathsf{SU}(2)_+ \times \mathsf{SU}(2)_- ~,~~~~~~~~I_3 \in \mathsf{SU}(2)_+~.
\end{align} 
These signs correspond to the signs in the second entry in \eqref{SUPDec}.
This means that the supercharges $Q_{\alpha J}$ in \eqref{SUPDec} have zero $I_3$ and imply that \eqref{Indz} is an index, independent of $\tilde{R}_9$. Similarly, the factor of $J^2$ in \eqref{Indz} absorbs the four fermion zero modes with $I_3=0$. These arise from the center-of-mass $\mathsf{U}(1)$ degrees of freedom.
These center-of-mass degrees of freedom have another four fermion zero modes that, together with the previous four, give the following contribution:
\begin{align}\label{Half}
{ B}_{2}(\nu)_{\rm half} = - {\rm Tr}[ (-1)^F \frac{(2 J)^2}{2}e^{ 2 \nu I_3} ] = e^{\nu} + e^{-\nu} -2 \equiv h(\nu) ~,~~~~~~~~h(\nu ) \sim \nu^2 ~{\rm as} ~\nu \to 0.
\end{align} 
 This contribution is present even for the two-point function, corresponding to a single D1-brane stretched between two D3-branes. The subscript ``half'' indicates that this is a half-BPS state, as opposed to the quarter-BPS states that describe the more generic situation. We always obtain the factor \eqref{Half} from the center-of-mass degrees of freedom. The computation of the rest is more elaborate, and we discuss it in detail in section~\ref{sec:IndexComputation}. For now, let us say that there are further zero modes that imply we need to insert more factors of $I_3$, as in \eqref{eq:IndexM}.
All these extra insertions are related to the insertions of supercharges in the ``closed string'' channel, which are related to the Grassmann variables in \eqref{AmplBFSS}. In other words, we claim that instead of \eqref{eq:review-open-closed} we should   consider   
 \begin{align} \label{CorrectRel}
\begin{array}{ll} 
{\cal I }_n \equiv &   -\frac{1}{[2(n-1)]!}\,{\rm Tr}\left[ (-1)^F \frac{(2J)^2}{2} (2 I_3)^{ 2 (n-1) } \right]  
\\ 
& \sim \langle B_1 | \langle B_2| \cdots \langle B_{n-1} | \prod_{i=1}^{n-1} ( w^{\alpha }_i Q^i_{\alpha K})^4  e^{ - (\tau_r - \tau_l)  (H_c - Z_c) } (\tilde{w}_n^{\dot \beta } \tilde{Q}^{n J}_{\dot \beta} )^4 |B_n \rangle  
\\
{\cal I}_n = & \mathcal{M}^{\rm BFSS}_n \mu^{2-n} ~.
\end{array} 
\end{align} 
The intermediate expression in terms of the supercharges is only schematic, and a more precise version will be described in the next subsection.

\subsection{More details on the connection between the index and the amplitude}

In the closed string channel, the insertions of $J$ and $I_3$ in \eqref{Indz} can be represented as integrals over $\tau$ of the corresponding rotation generators. These rotation generators couple to fermion zero modes in the closed string channel, $\chi_c$, which are responsible for generating the right state in the amplitude and are related to the factors of Grassmann variables in \eqref{AmplBFSS}. However, since these operators are integrated over $\tau$, they lead to factors of $\Delta\tau$. One purpose of this subsection is to show that such factors cancel against factors coming from bosonic light modes representing the center-of-mass motion of each external line. This cancellation was described in great detail for $n=3$ in~\cite{HM3}. Here we will be more schematic. We will focus on factors that are $n$-dependent, since we have already checked the case $n=3$.

The insertions of $J$ or $I_3$ involve ``open string'' channel fermion zero modes, $\chi_o$, which are related with the naturally normalized fermion zero modes of the closed string channel by~\cite{HM3}
\begin{equation}
\chi_o \mathrel{=} \sqrt{\frac{\Delta\tau}{2\pi\tilde R_9}}\chi_c
\end{equation}
where we are not distinguishing between $\Delta\tau_l$ and $\Delta\tau_l+\Delta\tau_r$, since this depends on the precise type of fermion zero mode~\cite{HM3}.  More precisely, the supercharge insertions are related to a canonically normalized fermion, $\{\chi_c,\chi_c^\dagger\}=1$, by $Q{=}{\frac{\sqrt{N_i}}{\sqrt{R_-}}}\chi_c$, so that $\{Q,Q^\dagger\}{=}{-}p_-{^{\,i}}{=}{\frac{N_i}{R_-}}$.
Here $Q$ denotes the canonically contracted component entering the insertion. This means that the insertions lead to a total factor of the form
 \begin{align}\label{GrassI}
J^2 I_3^{2(n-1)}  \to    \left(\frac{\Delta\tau R_-}{2\pi\widetilde R_9}\right)^{2n}\frac{1}{\prod_i^n N_i^2} \,\, \prod_{i=1}^{n-1}  (w.Q^i_K)^4  \times (\tilde w. \tilde{Q}^n_{J})^4
\end{align} 
where $J^2$ and $I_3$ are written in terms of $\chi_o$. We have also used that $w =(0,1)$, with no momentum dependence,  in the little group gauge \eqref{eq:ReviewSpinors} - \eqref{eq:GeneralMultiplet}. And the insertion of the supercharges takes into account all the factors involving Grassmann variables in \eqref{AmplBFSS}.
The upshot is that each factor contributes
\begin{equation}\label{eq:review-fermion-factor}
\delta^4(\eta_i^K)\ \longrightarrow\
\left(\frac{R_-}{N_i}\right)^2  \left(\frac{\Delta\tau}{2\pi\widetilde R_9}\right)^2 .
\end{equation}

Let us now discuss the bosonic zero modes. Each boundary state $|B_i\rangle$ produces the ground state of a $\mathsf{U}(N_i)$ $1+1$-dimensional gauge theory with electric flux $n_i$.
As shown in~\cite{WittenBoundStates}, this is a gapped state in the $\mathsf{SU}(N_i)$ theory for coprime $(N_i,n_i)$. Thus, we only need to consider the center-of-mass degrees of freedom within $\mathsf{U}(N_i)$, described by the $\mathsf{U}(1)_i$ degrees of freedom.
For the bosonic degrees of freedom, we have a zero-momentum state in the 123 directions and a definite position along the 4567 and 8 directions. We separate out the eighth direction because we will treat it differently. We can then write the boundary state as
\begin{equation}\label{BdySta}
|B_i \rangle = {\cal N}_i {\int\frac{d(-ip_8^i)\,d^4p^i}{(2\pi)^5}} |\vec p^{\, i}, p^i_8 \rangle ~, \qquad\qquad {\cal N}_i {=} {2\pi \sqrt{\frac{R_- \tilde{R}_9}{N_i}}}
\end{equation}
where ${\cal N}_i$ is a normalization factor~\cite{HM3}.
 These states propagate up to the interaction point and there they interact with an amplitude given by the ansatz 
 \eqref{AmplBFSS}, which is determined by the symmetries, up to the unknown factor $\mathcal{M}^{\textrm{BFSS}}$.

Let us first discuss the integrals over $p_8$. We see that the integrals over the $n-1$ factors of $p_8$ cancel against the explicit $p_8$-dependent delta functions in the amplitude without leaving any residual factor. The integral over the last $p_8$ can be viewed as an integral over the center-of-mass $\mathsf{U}(1)$ of the whole $\mathsf{U}(N)$ problem; it combines with other overall $\mathsf{U}(1)$ degrees of freedom, as discussed in~\cite{HM3}. Note that the condition $p_8^i=ip_9^i$ in \eqref{AmplBFSS} is only valid in the Lorentz frame where the second component of $\lambda^n$ is zero, or when $p_8^n=ip_9^n$. If that is not the case, then we should add to all particles the extra momentum $p_8^i=ip_9^i+zp_-^i$, where $z$ is an overall constant. Adding this corresponds to an overall small rotation of the string network in the $(\tau,8)$ plane. Of course, the rotated configuration continues to be BPS, but it preserves a different combination of supercharges, given by $\langle\lambda Q_I\rangle$ as indicated in \eqref{eq:PreservedSUSY}. When we discuss the integration over the overall $p_8$, we are referring to this constant $z$. Notice that, since the D3-branes are at definite positions, this integral is finite and only gives a factor of $\Delta\tau$. It is also present for the two- and three-point functions, $n=2,3$, as discussed in~\cite{HM3}.
 
Let us  discuss now the integral over the four momenta along 4567.  The $n-1$ left particles, and the final right particle lead to the integrals  
\begin{equation}\label{MonCo}
\int d^4 p_n e^{ - {\frac{R_-}{2}}\Delta \tau_r \vec p_n^{\, 2}/N } \prod_{i=1}^{n-1} d^4p_i e^{ - {\frac{R_-}{2}}\Delta \tau_l \vec p_i^{\, 2} /N_i } \delta^4\!\left( \sum_{i=1}^{n-1} \vec p_i - \vec p_n \right) \mathrel{=} {\frac{1}{4}\left(\frac{2\pi}{R_-}\right)^{2(n-1)}}\frac{1}{(\Delta\tau)^{2(n-1)}} \prod_{i=1}^{n-1} N_i^2
\end{equation}
where we set $\Delta \tau_l = \Delta \tau_r =\Delta \tau$. Note that the momentum conservation delta function comes from the amplitude. 

We see that this almost cancels all the factors of $\Delta\tau$ we had from \eqref{GrassI}. We are left with an $n$-independent factor of $\Delta\tau$ that cancels with the remaining momentum-conserving delta functions, as in the case $n=3$, the three-point amplitude; see~\cite{HM3}. This also involves factors coming from the open-string side due to the center-of-mass motion in the 123 directions: the index only counts the number of bound states, ignoring the center-of-mass motion.

So the final conclusion is that the factors of $\Delta\tau$ all cancel once we compare to the index, as expected. We also see below that the factors of $N_i$ cancel. This cancellation uses the fact that we have zero modes associated with each external line, something that is not completely obvious from the index computation that we describe below. This cancellation was indeed shown explicitly in~\cite{HM3}.

The overall factor (including $N$) is fixed by the $n$-independent two-point function and should work out because it works out for $n=3$. {Let us obtain the $n$-dependent prefactor explicitly. Substituting \eqref{eq:review-fermion-factor}, \eqref{BdySta}, and \eqref{MonCo} into \eqref{AmplBFSS} gives
\begin{equation}\label{FullZeroModeLedger}
\begin{aligned}
&\left[\frac{R_-\sqrt{R_9}}{(2\pi)^{n-9}\sqrt{N}}\kappa_{11}^{n-2}\right]
\prod_{i=1}^{n-1}\left[
\frac{{\cal N}_i}{(2\pi)^5}
\frac{2\pi \sqrt{N_i}}{R_-\sqrt{R_9}}
\left(\frac{\Delta\tau R_-}{2\pi\widetilde R_9N_i}\right)^2
\right]
\\[-1pt]
&\qquad\times
\left[\frac{1}{4}\left(\frac{2\pi}{R_-}\right)^{2(n-1)}
\frac{\prod_{i=1}^{n-1}N_i^2}{(\Delta\tau)^{2(n-1)}}\right] \longrightarrow \frac{\kappa_{11}^{n-2}}{\big[(2\pi)^4  (R_{-}R_9)^{1/2} \tilde{R}_9^{3/2}\big]^{n-2}} ,
\end{aligned}
\end{equation}
where we only kept the $n$-dependent piece. Using \eqref{eq:review-bfss-matrix-string-duality} and \eqref{eq:ReviewDictionary} gives
\begin{equation}\label{ClosedChannelFactor}
\left[
\frac{\kappa_{11}}{(2\pi)^4\ell_p^{3/2}\widetilde R_9}
\right]^{n-2}
=\left(\frac{R_-R_9}{\sqrt{2}}\right)^{n-2}
=\mu^{2-n}.
\end{equation}
}
 The conclusion is that we have established that 
 \begin{align}
 { \cal I }_n =  -\frac{1}{[2(n-1)]!}\,{\rm Tr}\left[ (-1)^F \frac{(2J)^2}{2}  (2 I_3)^{  2(n-1) }\right]=  \mu^{2-n}\mathcal{M}_n^{\rm BFSS}\,.
\end{align} 
 We stress that these arguments implying the relation between the BFSS amplitude and the index do not depend on a kinematical chamber (i.e. whether the momenta obey \eqref{eq:later-decay-chamber}  or not). In general, however, the index may be hard to evaluate; we see below that a simple expression can be found in the  decay chamber \eqref{eq:later-decay-chamber}.

\section{Computing the index }
\label{sec:IndexComputation}

In this section, we review the method that Sen developed for the computation of the index~\cite{Sen}, and we apply it to the case of interest.
As we mentioned previously, it is convenient to compute the following index first, defined  in~\cite{Sen}:
\begin{align} \label{B2Index}
B_2(\nu) =- {\rm Tr}[ (-1)^F \frac{(2J)^2}{2} e^{ \nu 2 I_3 }]~.
\end{align} 
This is an index because there is a supercharge that has $I_3=0$. 

This problem was considered in~\cite{Sen} from the point of view of string networks stretching between D3-branes. In principle, our problem is slightly different because it involves the $1+1$-dimensional super Yang--Mills theory living on the D1-branes. However, we expect that the answer should be the same because we can start from the original string-theory configuration and take the field-theory limit. Accordingly, we assume here  that the wall-crossing methods discussed below apply directly to the two-dimensional field-theory problem. A proof would be of interest.

The problem with the D3-branes has one interesting feature: the $SL(2,\mathbb Z)$ symmetry under which the $(N,n)$ charges transform as a doublet. Invariance of the index under $SL(2,\mathbb Z)$ implies that $\mathcal{M}^{\rm BFSS}$ is Lorentz invariant in the eleven-dimensional sense once we make $N_i,n_i$ continuous.
We can view the problem involving the D3-branes and the resulting string networks as a problem in four-dimensional ${\cal N}=4$ super Yang--Mills theory with gauge group $\mathsf{U}(n)$, where we go to the Coulomb branch by separating all the D3-branes. This breaks $\mathsf{U}(n)\to\prod_i \mathsf{U}(1)_i$. In that theory, we are computing degeneracies of BPS states. These states have magnetic and electric charges $N_i$ and $n_i$, respectively, under each $\mathsf{U}(1)_i$. From this point of view, it is natural to define the vector $\Gamma_i$ introduced in
\eqref{eq:GammaDefinition}. The $SL(2,\mathbb Z)$ symmetry is the usual electric-magnetic duality symmetry of ${\cal N}=4$ SYM.

The degeneracies of the BPS states do not depend on small changes in the positions of the D3-branes, at least while the positions remain within a chamber. However, there are walls of marginal stability, and the spectrum can change when we cross them. For our problem, we are interested in computing the index in the region where the D3-brane positions are as indicated in figure \ref{fig:Scattering}. One can conquer the index by scaling the walls of marginal stability. For this, we use the ``wall-crossing formula''~\cite{Sen,DenefMoore,SenBlack,KS}.
This formula tells us how the spectrum, and therefore the index, changes under crossing a wall. Therefore we can start from a region where the index is simple, and then cross several walls to get to the point of interest. Sen explained how to perform this computation.

\subsection{A first example: the three-point function}

\begin{figure}[htbp]
   \begin{center}
   \includegraphics[scale=.4]{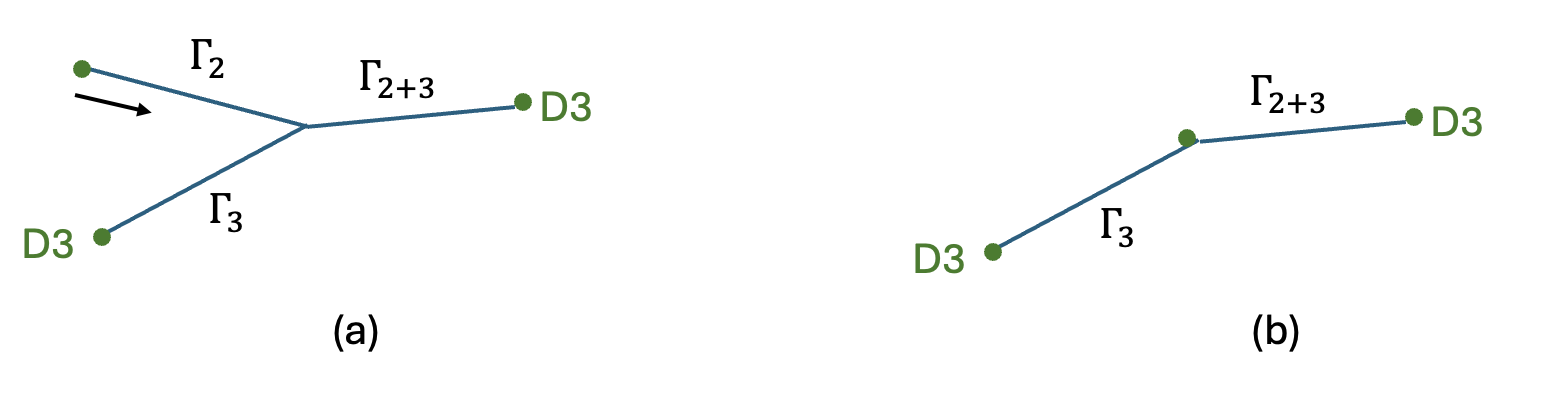}
    \end{center}
    \caption{(a) The starting configuration of brane positions. We move the D3-brane attached to the line labeled by $\Gamma_2$ toward the intersection point. (b) When the brane reaches the intersection point, we encounter a wall of marginal stability where the BPS state can decay into the two separate lines indicated in the figure.}
    \label{ThreeMarginal}
\end{figure}

As a first example, we consider the three-point function. For later use, we label the three external legs as $2,3$, and $2+3$; see figure \ref{ThreeMarginal}.

As we move the D3-brane position of particle 2 downward, we reach a wall of marginal stability when the intersection point reaches particle 3. There, the bound state decays into two half-BPS states, and the combined BPS state ceases to exist.


Across a primitive wall with positive intersection $m$, Sen's formula~\cite{Sen}
multiplies the product of the two constituent indices by $m$, up to the
standard protected-spin sign. Here ``primitive'' means that the two decay products have charge vectors $\Gamma$ that cannot be written as integer multiples of other charge vectors. In this case, this is ensured if all $(N_i,n_i)$ are coprime.
Thus, for an ordered decay
wall with $m:=-\langle\alpha,\beta\rangle>0$, the primitive rule for the discontinuity in the index is\footnote{Here we must also constrain the charges so that the composite $\Gamma_2+\Gamma_3$ sector is itself primitive, isolated, and gapped. The first assumption trivializes an extra sign $(-1)^{\langle \alpha, \beta\rangle}$ in the wall-crossing formula~\cite{Sen}.}
\begin{align}
  \Delta B_2(\alpha+\beta;\nu)
  =
  m\,
  B_2(\alpha;\nu)
  B_2(\beta;\nu).
  \label{eq:five-primitive-wall}
\end{align}
On the side of the first wall, where the $(23)$ bound state is absent, the 
index vanishes.  Equation~\eqref{eq:five-primitive-wall} therefore gives,
on the other side,
\begin{align}  
  B_2(\nu)
  =
  h(\nu)^2m_2 ~,~~~~~~~~~m_2:=
  -\langle\Gamma_3,\Gamma_2\rangle
  =
  \frac{1}{\mu}[32]>0.
  \label{eq:five-first-binding}
\end{align}
where we used the half-BPS index given in \eqref{Half}.
Expanding this to order $\nu^4$,  we get 
\begin{align}
B_2(\nu) \sim \nu^4 {\cal I}_3  ~,~~{\rm as } ~\nu \to 0 ~,~~~~~~\longrightarrow ~~~~~~ {\cal I}_3 =  -\langle\Gamma_3,\Gamma_2\rangle = \frac{1}{\mu} \mathcal{M}^{\rm BFSS}_3~.
\end{align}

\subsection{Facing the faces}

\label{sec:Faces}

\begin{figure}[htbp]
   \begin{center}
   \includegraphics[scale=.4]{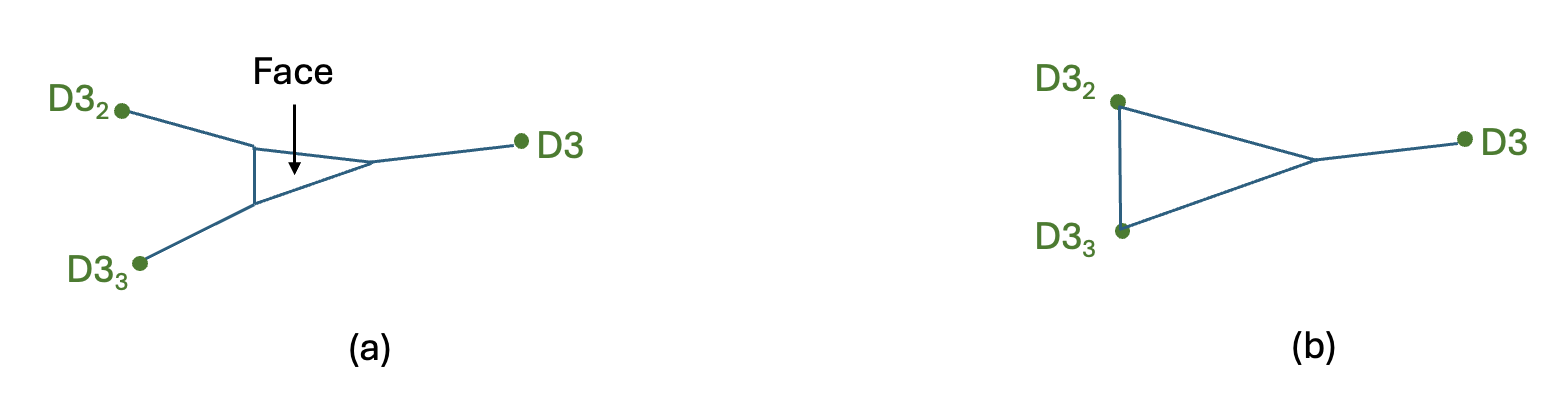}
    \end{center}
    \caption{ (a)  A configuration with a face. We could further break each of the vertices into a subface. For large $(N_i,n_i)$ we can have many faces and subfaces. 
   (b) A wall of marginal stability associated to the face in (a).    }
    \label{Faces}
\end{figure}

Actually, the computation in \eqref{eq:five-first-binding} does not give the full $\nu$ dependent index. In this section we will discuss a feature that is  irrelevant for the computation of ${\cal I}_3$, but is important for  the computation of $B_2(\nu)$. The string network diagram can have faces, as displayed in figure \ref{Faces}.  

When a face becomes large, it can touch one of the D3-branes. If it touches only one D3-brane, that is not an issue: it is not a wall because the network cannot decay into two disconnected states. However, if a face touches two D3-branes, as happens in figure \ref{Faces}(b), then we have a wall where the configuration decays into a half-BPS state, the string between D3$_2$ and D3$_3$, and a three-pronged string connecting all the D3-branes. Notice that this wall occurs when branes two and three are equidistant. In particular, it can occur even if both branes are far away; it depends only on the ordering of their distances. In this case, the wall-crossing formula gives a factor of $h(\nu)$ for the half-BPS object and a factor of the form $B_2(\nu)$ for the remaining three-pronged string. An explicit example was discussed in~\cite{Sen}. However, the jump at this wall crossing is proportional to $\nu^6$ for small $\nu$. We conclude that it does not affect the index ${\cal I}_3$, though it certainly affects $B_2(\nu)$. The basic reason for this extra power of $\nu$ is that one of the objects into which the network decays still has the same connectivity properties as the original network. This is a general property of faces, even when they are present in a string network with more external states.

In principle, there can be a complicated pattern of faces and subfaces. For example, each of the vertices in figure \ref{Faces}(a) could be split into further triangles. Such faces are easy to produce when $(N_i,n_i)$ are large, though they are not possible for very small $(N_i,n_i)$. Fortunately, this is a complication that we do not have to worry about in this paper. It might be interesting to know whether there is some computation in M-theory where such faces are important. Note that they look like loop diagrams, so they might be loop corrections to some computation.

\subsection{The four-point function}

\begin{figure}[htbp]
   \begin{center}
   \includegraphics[scale=.4]{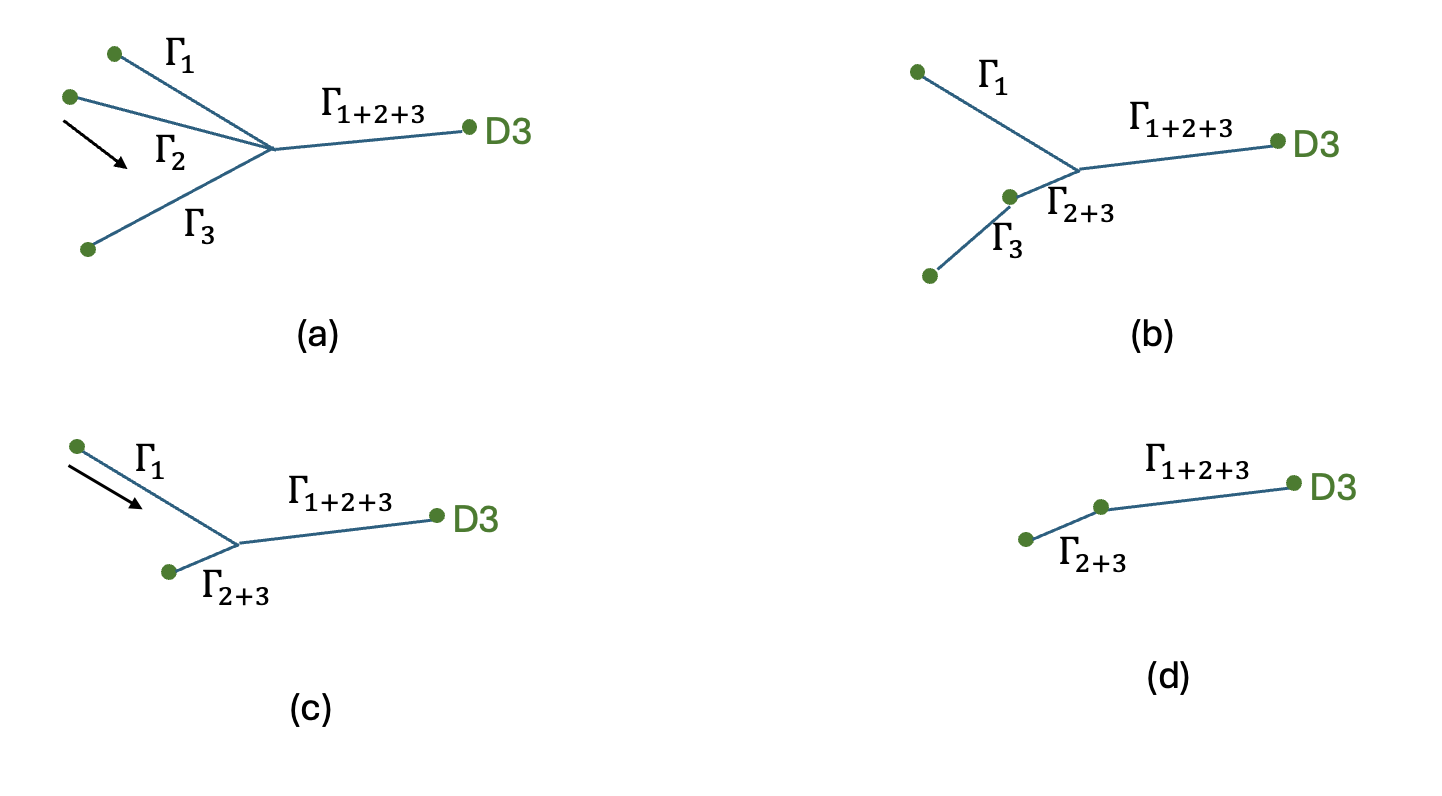}
    \end{center}
    \caption{(a) The starting configuration of brane positions. We move the D3-brane attached to the line labeled by $\Gamma_2$ toward the intersection point and slightly downward to bring out the line $2+3$. (b) When the brane reaches the intersection point, we encounter a wall of marginal stability.
    (c) We focus on one of the decay products of (b). Then we move the D3-brane attached to line 1 to the intersection point and again use the same formula as in figure \ref{ThreeMarginal}.}
    \label{FourMarginal}
\end{figure}

For the four-point function, we sequentially apply Sen's formula. First, we move the D3-brane of particle 2 as in figure \ref{FourMarginal}. We move it downward slightly to resolve the vertex and bring out the line $2+3$. This rotates the network slightly while preserving all relative angles, so that a slightly different supercharge is preserved. This small rotation is irrelevant for the purposes of our discussion. We then move D3$_2$ to the interaction point and reach the wall of marginal stability; see figure \ref{FourMarginal}(b). We apply the wall-crossing formula to obtain
\begin{align}
B_2(\nu)_{1,2,3,4}  = m_2 h(\nu) B_2(\nu)_{(1,2+3,4) }
\end{align} 
with $m_2$ given in \eqref{eq:five-first-binding}. 

Focusing now on the decay product labelled by the external charges $\Gamma_1,\Gamma_2 +\Gamma_3 , \Gamma_4$ we then move the D3 brane attached to the line 1; see figure \ref{FourMarginal}(c). 
At the wall of marginal stability, figure \ref{FourMarginal}(c), we apply a formula similar to \eqref{eq:five-first-binding}.
We get 
\begin{align}
B_2(\nu)_{1,2,3,4}  = m_2 m_1 h(\nu)^3  
\end{align}
 where we used that the second intersection number 
  between $\Gamma_1$ and the charge
already formed at the first vertex is 
\begin{align}
  m_1:=
  -\langle\Gamma_2+\Gamma_3,
          \Gamma_1\rangle~.
  \label{eq:five-m1}
\end{align}
Again, expanding to order $\nu^6$ we get 
\begin{align}\label{Fin4Ind}
 {\cal I}_4 = \langle \Gamma_3,\Gamma_2 \rangle \langle \Gamma_2 + \Gamma_3 , \Gamma_1 \rangle  = \frac{1}{\mu^2} [32]( [21]+ [31])= \frac{1}{\mu^2}\mathcal{M}^{\rm BFSS}_4~.
\end{align} 
The equalities on the right follow directly from
\eqref{eq:ReviewDictionary}.

As a side comment, note that if we had initially moved the D3$_2$
brane slightly upward from the starting configuration in figure \ref{FourMarginal}, then we would first have joined lines 1 and 2 into a line with charges $1+2$, which would then join line 3. In this case, it would be {\it wrong} to calculate the index as a product of the three-point index for each vertex. The reason is that, if we were to move D3$_2$ to the new interaction vertex, then on the other side of that wall we could still have a BPS state involving a fully connected network. Taking this into account gives the correct formula, \eqref{Fin4Ind}. The point of this comment is to emphasize that the computation of the index is not simply a matter of multiplying the index for each three-point vertex in a string network. It reduces to that only in special cases. In this paper, we choose the external kinematics to be in the special chamber \eqref{eq:IntroDecayChamber}, which ensures that this happens when we write the network appropriately.

\subsection{General \texorpdfstring{$n$}{n}}

\begin{figure}[htbp]
   \begin{center}
   \includegraphics[scale=.4]{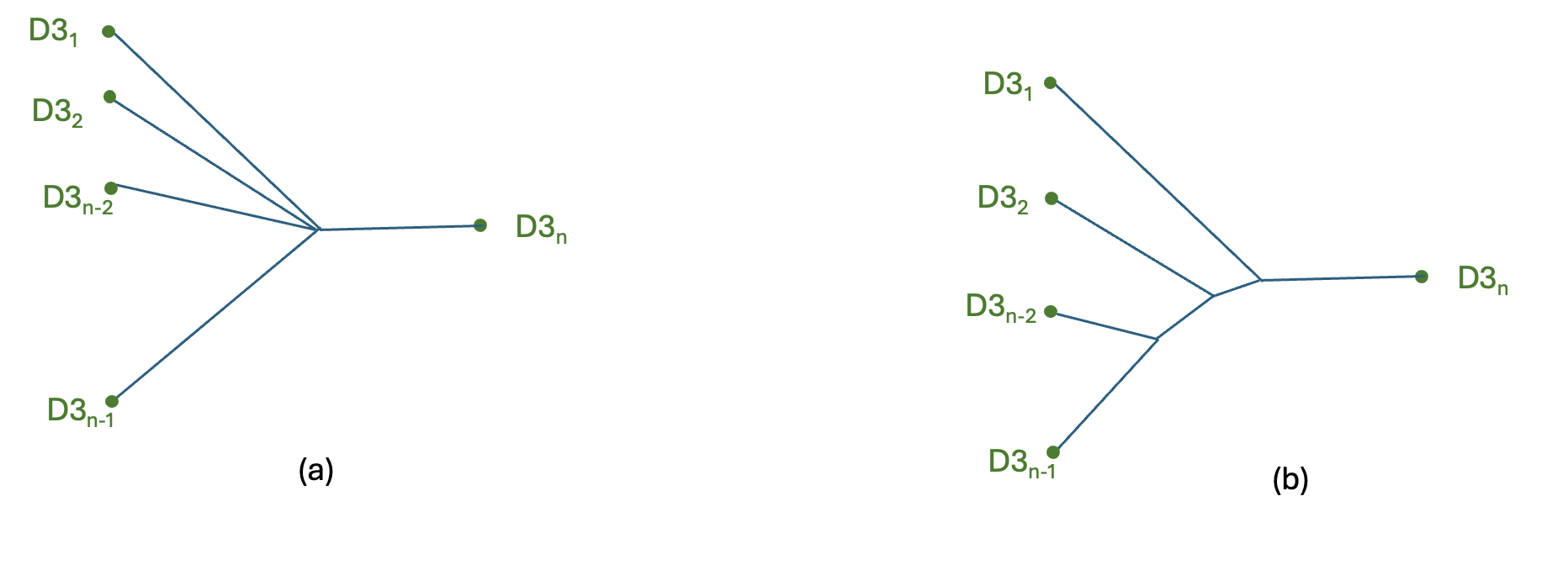}
    \end{center}
    \caption{ (a) The $n$ point problem in the chamber \eqref{eq:later-decay-chamber} for the external states. (b) We deform the position of the D3 branes   to an ordered comb as depicted. Then the computation can be done by sequentially bringing the D3 branes to the interaction vertices, starting from D3$_{n-2}$ and working our way up all the way to D3$_1$.     }
    \label{nPoint}
\end{figure}

We now consider the general-$n$ case in the special chamber \eqref{eq:IntroDecayChamber} for the external states. This implies a condition on the values of $n_i/N_i$ 
\begin{align}
 \frac{n_1}{N_1} < \cdots < \frac{n_{n-2}}{N_{n-2}} < \frac{n_n}{N_n} < \frac{n_{n-1}}{N_{n-1} },
  \label{eq:later-decay-chamber}
\end{align}
which becomes a condition for the slopes of the lines; see figure \ref{nPoint}.

Following a similar procedure, as indicated in figure \ref{nPoint}, we get 
\begin{align}\label{IndexFin}
B_2(\nu) = h(\nu)^{n-1} \prod_{a=1}^{n-2} m_a    
\end{align} 
where 
\begin{align}
  \Gamma_{>a}:=\sum_{b=a+1}^{n-1}\Gamma_b,
  \qquad
  m_a:=-\langle\Gamma_{>a},\Gamma_a\rangle
  =
  \frac{1}{\mu}
  \sum_{b=a+1}^{n-1}[ba]>0 .
  \label{eq:six-multiplicities}
\end{align}

We see that \eqref{IndexFin} indeed has a zero of order $\nu^{2(n-1)}$. We should emphasize that \eqref{IndexFin} was derived by repeated application of the wall-crossing formula~\cite{SenBlack}.
This implies that 
\begin{align}\label{FinalNpt}
{\cal I }_n = \prod_{a=1}^{n-2} m_a =  \mu^{2-n}\mathcal{M}^{\rm BFSS}_n|_{\mathcal{R}_{n,n-1}}  
\end{align} 
with $m_a$ as in \eqref{eq:six-multiplicities}. This recovers \eqref{eq:IntroProduct}.

\section{Comparing to the gravity amplitude}
\label{sec:Comparison}
 
As we mentioned above, the gravity amplitude is written as a degree-$n-2$ polynomial in the $[ij]$ brackets~\cite{GLSSW}. The precise polynomial changes as we cross the boundaries of certain chambers. A simple chamber is the so-called decay chamber, where the answer factorizes and takes the form
\eqref{eq:IntroProduct}. In this chamber, we have checked the precise agreement with the index once we realize that the brackets $[ij] $ are proportional to $\langle \Gamma_i , \Gamma_j\rangle $. Notice that in the supergravity theory, the single-minus amplitude cannot have any quantum correction. The reason is that any quantum correction to the supersymmetric amplitude \eqref{eq:SUSA}  should involve powers of $G_N (p^i.p^j)^{9/2}$,   but supersymmetry implies that all invariants are zero, $p^i.p^j=0$.   Of course, in non-supersymmetric theories we can have single-minus amplitudes that are non-zero for general kinematics. 

Notice that the agreement is complete even at finite values of $N_i$ and $n_i$, which correspond to the naive compactification of the gravity theory. In principle, the BFSS conjecture only requires agreement for large $N_i,n_i$.

Our calculation keeps the numerical factors and the factors of $l_{\rm Planck}$, $R_-$, and $R_9$.  It imports the canonically normalized two-point and boundary-state inputs checked in~\cite{HM3} and the stated open/closed-channel identification.  Under these explicit assumptions, the resulting agreement is exact in the stated chamber.

Let us make some comments 

\begin{itemize}
 
\item 
We expect that the connection between the index and the single-minus amplitude should continue to hold in all chambers for the external momenta. Here we consider only a particularly simple one. In particular, we also expect it to hold when we have any number of initial and any number of final states. It would be nice to check this. 

\item
We have also considered only primitive vectors for the external lines and all possible internal lines entering the diagram. It would be nice to explore whether anything interesting happens when this is not the case. From the point of view of BFSS where $(N,n)$ become large and continuous, it looks like we can always take the large $N$ limit in such a way that we have coprime values. However, one would still like to understand whether there is some interesting correction when they are not coprime. Any such correction should become unimportant, or subleading, as we take the large $N $ limit.  Verifying this would be a nice test of the BFSS conjecture (and our method to connect the amplitude to the index). 

\item 

It seems that our argument connecting the amplitude and the index could be simplified further. We noticed several cancellations between some of the bosonic transverse modes and some of the fermionic modes.

\item  One can also ask whether the full index $B(\nu)$ corresponds to an M-theory computation. It is clear from the definition of the index \eqref{B2Index} that we can think of $\nu$ as an additional twist that we perform when we compactify the coordinate $\tilde x^9$. Namely, when we identify $\tilde x^9 \to \tilde x^9 + 2 \pi \tilde R_9 $ we also perform a rotation in four of the coordinates orthogonal to the D3 branes as $ z_i \to e^{ i \theta/2}  z_i $, where $i=1,2$ are two complex coordinates. We identify $2 \nu = i \theta$. This is sometimes called a ``Melvin" background. We can now T-dualize back to the original $x^9$ coordinate to find a background with an NS $B$ field. This can then be lifted to eleven dimensions. In this way, we can view the problem as computing a scattering amplitude in a non-trivial background.  Moreover, Sen also introduced another index that depends on an additional parameter $B(\nu,y)$ where $y$ is related to a chemical potential for $J+I_3^-$~\cite{Sen}. This can also be lifted to M-theory in a similar fashion.  It would be interesting to study both of these problems in more detail.

\item 

It seems interesting that the computation of the index is related to a type of Feynman diagram, or, more precisely, a Landau-type diagram~\cite{GLSSW}. Could this give rise to a simpler way to compute the index?

\item 

It would be nice to understand further the recursion relations that compute the amplitude and to relate them explicitly to the ones that compute the index. 

\item 
Notice that the walls that we have been crossing to compute the index in section~\ref{sec:IndexComputation} separate chambers in the space of positions for the D3 branes.  
In crossing these walls, we keep the external charges $(N_i,n_i)$ fixed, and hence the external momenta fixed. The amplitude is the index in a particular chamber where the D3-branes are arranged as in figure \ref{fig:Scattering}. The motion of the D3-branes and their wall crossing are just a trick to compute the index.

On the other hand, there are also chambers in the space of external charges $(N_i,n_i)$. These chambers are clearly seen in the gravity computation~\cite{GLSSW}, where the form of the formula changes as we change the external kinematics. In other words, if we take $(N_i,n_i)$ to be continuous and vary them, then we can encounter walls where the form of $\mathcal{M}^{\rm grav}$ changes.  For example, for the very simple case of $n=3$, the gravity result is 
\begin{align}
\mathcal{M}_3^{\rm grav} \propto \left| [ 12] \right| 
\end{align} 
We see that there are two kinematic chambers, set by the sign of $[12]$.  As we increase $n$ we have a more complicated pattern of chambers.  Within a chamber, $\mathcal{M}^{\rm grav}_n$ is a polynomial of degree $n-2$ in $[ij]$.
  This polynomial  may change at loci   where~\cite{GLSSW} 
\begin{align}\sum_{i\in S} \sum_{j\in T}[ij]=0 \end{align} for subsets $S,T$ of the particles,  with $S\cup T$ giving $n-1$ particles. 
This can also be viewed as a wall-crossing phenomenon, but it is different from  the wall crossing involved in the calculations in section~\ref{sec:IndexComputation}. The gravity wall crossing arises when we change the external kinematics, or equivalently the $(N_i,n_i)$. Such changes lead to different expressions for the index in terms of the $\langle\Gamma_i,\Gamma_j\rangle$ brackets.

\item As we outline in the next section, inside a given kinematic chamber the form of the amplitude is completely determined by $w_{1+\infty}$ soft theorems~\cite{GLSSW}. Since tree Feynman diagrams realize the loop group ${\cal L}w_{1+\infty}$, one expects the gravity wall crossings to represent ${\cal L}w_{1+\infty}$ (this has not been shown). On the other hand, de~Wit, Hoppe, and Nicolai~\cite{DHN} showed that the large $N$ limit of the matrix model has a  $w_{1+\infty}$ symmetry, which becomes ${\cal L}w_{1+\infty}$ in the 1+1 matrix theory. It would be interesting to understand these statements and their possible connections better.

\end{itemize}

\def\be{\begin{equation}}
\def\ee{\end{equation}}
 \def\in{{\rm in}}
\def\out{{\rm out}}

\section{$w_{1+\infty}$ soft theorems from wall crossing}
\label{sec:Soft}

The rest of the paper is devoted to further checks and implications of the amplitude-index equivalence. In this section, we derive the \(w_{1+\infty}\) tower of soft theorems~\cite{Guevara:2021abz,Strominger:2021mtt} for single-minus BFSS amplitudes directly from wall crossing in D3-brane position space.

We consider a single-minus ($n=n_\in+n_\out+1$)-point scattering amplitude with momenta obeying
\be
 \sum_{j=1}^{n_\out}|j]_\out+ \sum_{j=n_\out+1}^{n_\out+n_\in}|j]_\in
 +|s]=0 .
\ee
where $|s]$ is taken to be the incoming positive-helicity `soft' particle,  and we refer to the other particles as `hard'. By the particle being soft, we mean that its momentum is much smaller than that of the other particles
\begin{equation} \label{SoftCond}
|n_s| , |N_s | \ll |n_i| , |N_i |
\end{equation}
Outgoing hard legs are labeled  by $1,\ldots,n_\out$ and the incoming ones  by $n_\out+1,\ldots,n_\out+n_\in$. We arrange the positions of the D3 branes  associated to the hard particles so that the corresponding lines meet in a central interaction point, as in Figure~1. 

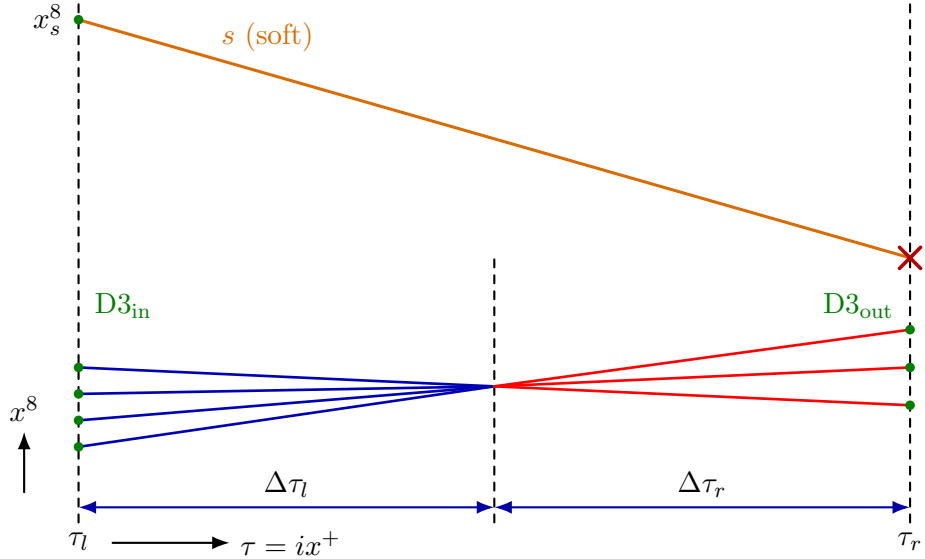
\begin{figure}[htbp]
  \centering
  \begin{tikzpicture}[
      x=1cm,
      y=1cm,
      line cap=round,
      line join=round,
      >=Latex
    ]
    \coordinate (hard) at (5.5,2.35);

    \draw[dashed, line width=0.8pt] (0,0.55) -- (0,7.50);
    \draw[dashed, line width=0.8pt] (5.5,0.55) -- (5.5,4.05);
    \draw[dashed, line width=0.8pt] (11,0.55) -- (11,7.50);
    \draw[blue!65!black, <->, line width=0.8pt]
      (0,0.75) -- node[above, text=black] {$\Delta\tau_l$} (5.5,0.75);
    \draw[blue!65!black, <->, line width=0.8pt]
      (5.5,0.75) -- node[above, text=black] {$\Delta\tau_r$} (11,0.75);
    \node[below] at (0,0.55) {$\tau_l$};
    \node[below] at (11,0.55) {$\tau_r$};

    \foreach \yy in {1.55,1.90,2.25,2.60} {
      \draw[blue!70!black, line width=1.0pt]
        (0,\yy) -- (hard);
      \fill[green!50!black] (0,\yy) circle[radius=1.8pt];
    }
    \foreach \yy in {2.10,2.60,3.10} {
      \draw[red, line width=1.0pt]
        (11,\yy) -- (hard);
      \fill[green!50!black] (11,\yy) circle[radius=1.8pt];
    }

    \draw[orange!85!black, line width=1.15pt]
      (0,7.20) -- (11,4.05);
    \fill[green!50!black] (0,7.20) circle[radius=1.8pt];
    \draw[red!70!black, line width=1.35pt]
      (10.86,3.91) -- (11.14,4.19)
      (10.86,4.19) -- (11.14,3.91);
    \node[above, orange!85!black] at (2.5,6.65) {$s\ \text{(soft)}$};
    \node[left] at (-0.08,7.20) {$x_s^8$};

    \node[green!50!black, right] at (0.08,3.42) {$\mathrm{D3}_{\rm in}$};
    \node[green!50!black, left] at (10.92,3.42) {$\mathrm{D3}_{\rm out}$};
    \draw[->, line width=0.8pt] (-0.72,0.95) -- (-0.72,1.75)
      node[above] {$x^8$};
    \draw[->, line width=0.8pt] (0.45,0.28) -- (2.0,0.28)
      node[right] {$\tau=ix^+$};
  \end{tikzpicture}
  \caption{The index $\widetilde {\cal I} _n|_{\rm fig.\,\ref{fig:wsoft-vanishing}}=0$. The hard strings meet at a central interaction point, as in Figure~\ref{fig:Scattering}. The soft string has boundary value $x_s^8\gg\Delta\tau$ and the slope fixed by $|s]$, so it misses the hard web and remains disconnected; the cross at $\tau_r$ marks its forbidden boundary endpoint.}  \label{fig:wsoft-vanishing}
\end{figure}

Let us now take
the boundary value of the position for the D3 brane associated to the insertion point of the soft particle, $x^8_s$,  to be much greater than $\Delta\tau$ so that, with the angle specified by $|s]$, it cannot meet the string web associated with the hard particles. In addition, we put a restriction on the angle for the soft particle, set by $n_s/N_s$. We require that it is larger than the slopes of all other particles. More precisely, we impose that  $[sj]>0$ for $j$ outgoing and $[sj]<0$ for $j$ incoming. 
This angle restriction makes the following visualization slightly easier but is not required for the final conclusion, as we will explain later.
The index for this configuration vanishes because there is no place where the string associated to the soft particle can end.  See figure~\ref{fig:wsoft-vanishing}. Note that, due to \eqref{SoftCond}, the addition of the rest of the particles determine which supersymmetry is preserved and therefore the slope associated to the soft particle. 
Moreover, the lower subweb vanishes  because it does not conserve overall momentum.

\FloatBarrier

\begin{figure}[htbp]
  \centering
  \begin{tikzpicture}[
      x=1cm,
      y=1cm,
      line cap=round,
      line join=round,
      >=Latex
    ]
    \coordinate (hard) at (5.5,2.35);
    \coordinate (softjunction) at (10.38,3.16);

    \draw[dashed, line width=0.8pt] (0,0.55) -- (0,6.45);
    \draw[dashed, line width=0.8pt] (5.5,0.55) -- (5.5,4.05);
    \draw[dashed, line width=0.8pt] (11,0.55) -- (11,6.45);
    \draw[blue!65!black, <->, line width=0.8pt]
      (0,0.75) -- node[above, text=black] {$\Delta\tau_l$} (5.5,0.75);
    \draw[blue!65!black, <->, line width=0.8pt]
      (5.5,0.75) -- node[above, text=black] {$\Delta\tau_r$} (11,0.75);
    \node[below] at (0,0.55) {$\tau_l$};
    \node[below] at (11,0.55) {$\tau_r$};

    \foreach \yy in {1.55,1.90,2.25,2.60} {
      \draw[blue!70!black, line width=1.0pt]
        (0,\yy) -- (hard);
      \fill[green!50!black] (0,\yy) circle[radius=1.8pt];
    }
    \foreach \yy in {2.10,2.60} {
      \draw[red, line width=1.0pt]
        (11,\yy) -- (hard);
      \fill[green!50!black] (11,\yy) circle[radius=1.8pt];
    }

    \draw[violet!75!black, line width=1.1pt] (hard) -- (softjunction);
    \draw[orange!85!black, line width=1.15pt]
      (0,6.05) -- (softjunction);
    \draw[red, line width=1.0pt] (softjunction) -- (11,3.10);
    \fill[green!50!black] (0,6.05) circle[radius=1.8pt];
    \fill[green!50!black] (11,3.10) circle[radius=1.8pt];

    \node[above, orange!85!black] at (2.45,5.42) {$s\ \text{(soft)}$};
    \node[left] at (-0.08,6.05) {$x_s^8$};
    \node[above, violet!75!black] at (7.30,2.82) {$1+s$};
    \node[red, anchor=west] at (11.12,3.10) {$1$};

    \draw[->, black, line width=0.8pt]
      (10.92,3.23) -- (10.62,3.26);

    \node[green!50!black, right] at (0.08,3.42) {$\mathrm{D3}_{\rm in}$};
    \node[green!50!black, left] at (10.92,3.82) {$\mathrm{D3}_{\rm out}$};
    \draw[->, line width=0.8pt] (-0.72,0.95) -- (-0.72,1.75)
      node[above] {$x^8$};
    \draw[->, line width=0.8pt] (0.45,0.28) -- (2.0,0.28)
      node[right] {$\tau=ix^+$};
  \end{tikzpicture}
  \caption{
  $\widetilde {\cal I}_n|_{\rm fig. \ref{fig:wsoft-crossing}}(1,\ldots,n-1,s)=\langle \Gamma_1,\Gamma_s\rangle {\cal I}_{n-1}(1+s,2,\ldots,n-1)$.
  After $x_s^8$ is lowered past the uppermost outgoing leg, the soft string $s$ binds to leg $1$ at a trivalent junction. The composite leg $1+s$ joins the remaining hard strings at the central interaction point. The arrow indicates the displacement direction of the uppermost late-time D3 brane used in the wall-crossing computation.}
  \label{fig:wsoft-crossing}
\end{figure}
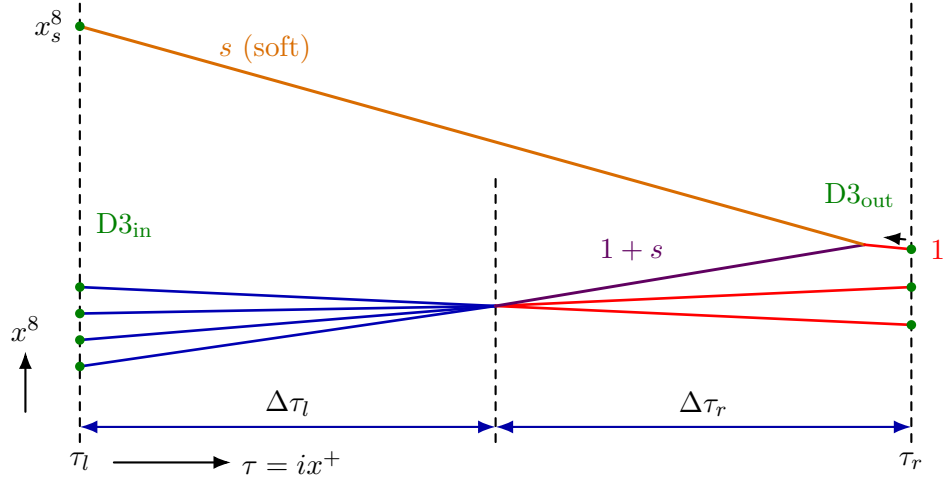

Now let's move $x_s^8$ slowly downwards until just after the soft string crosses the uppermost outgoing string, numbered ``1''.   Let's also slightly adjust $x^8_1$ so that the $1+s$ string   hits the interaction point. This is depicted in Figure~\ref{fig:wsoft-crossing}. The resulting momentum-conserving string web can be computed by moving the uppermost late-time D3 brane slightly in until it crosses the junction with the soft string and the amplitude vanishes.  Primitive wall crossing then gives
\be \label{FirstCross}
  \widetilde {\cal I}_n|_{\rm fig. \ref{fig:wsoft-crossing}}(1,\ldots,n-1,s)
 =\langle \Gamma_1 , \Gamma_s \rangle \,{\cal I }_{n-1}(1+s,2,\ldots,n-1).
\ee

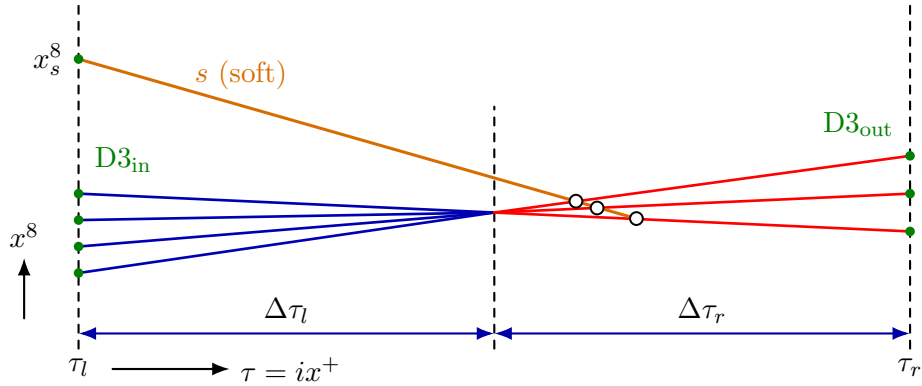
\begin{figure}[htbp]
  \centering
  \begin{tikzpicture}[
      x=1cm,
      y=1cm,
      line cap=round,
      line join=round,
      >=Latex
    ]
    \coordinate (hard) at (5.5,2.35);

    \draw[dashed, line width=0.8pt] (0,0.55) -- (0,5.15);
    \draw[dashed, line width=0.8pt] (5.5,0.55) -- (5.5,3.75);
    \draw[dashed, line width=0.8pt] (11,0.55) -- (11,5.15);
    \draw[blue!65!black, <->, line width=0.8pt]
      (0,0.75) -- node[above, text=black] {$\Delta\tau_l$} (5.5,0.75);
    \draw[blue!65!black, <->, line width=0.8pt]
      (5.5,0.75) -- node[above, text=black] {$\Delta\tau_r$} (11,0.75);
    \node[below] at (0,0.55) {$\tau_l$};
    \node[below] at (11,0.55) {$\tau_r$};

    \foreach \yy in {1.55,1.90,2.25,2.60} {
      \draw[blue!70!black, line width=1.0pt] (0,\yy) -- (hard);
      \fill[green!50!black] (0,\yy) circle[radius=1.8pt];
    }
    \foreach \yy in {2.10,2.60,3.10} {
      \draw[red, line width=1.0pt] (hard) -- (11,\yy);
      \fill[green!50!black] (11,\yy) circle[radius=1.8pt];
    }

    \draw[orange!85!black, line width=1.15pt] (0,4.38) -- (7.38,2.27);
    \fill[green!50!black] (0,4.38) circle[radius=1.8pt];
    \node[left] at (-0.08,4.38) {$x_s^8$};
    \node[above, orange!85!black] at (2.15,3.84) {$s\ \text{(soft)}$};

    \foreach \x/\y in {6.58/2.50,6.86/2.41,7.38/2.27} {
      \fill[white] (\x,\y) circle[radius=2.4pt];
      \draw[black, line width=0.75pt] (\x,\y) circle[radius=2.4pt];
    }

    \node[green!50!black, right] at (0.08,3.05) {$\mathrm{D3}_{\rm in}$};
    \node[green!50!black, left] at (10.92,3.52) {$\mathrm{D3}_{\rm out}$};
    \draw[->, line width=0.8pt] (-0.72,0.95) -- (-0.72,1.75)
      node[above] {$x^8$};
    \draw[->, line width=0.8pt] (0.45,0.28) -- (2.0,0.28)
      node[right] {$\tau=ix^+$};
  \end{tikzpicture}
  \caption{
  $\widetilde {\cal I}_n|_{\rm fig.\,\ref{fig:wsoft-outgoing-sum}}(1,\ldots,n-1,s) =\sum_{j=1}^{n_{\rm out}}\langle \Gamma_j , \Gamma_s\rangle \,{\cal I}_{n-1}(1,\ldots,j+s,\ldots,n-1)$.
  With the soft trajectory just above the central interaction point, it crosses all $n_{\rm out}$ outgoing legs. Each open circle denotes one possible trivalent junction where the soft string may end and hence one term in the sum over outgoing strings.}
  \label{fig:wsoft-outgoing-sum}
\end{figure}

\FloatBarrier
Now let us continue decreasing $x_s^8$ so that the soft string passes just above the central interaction point, as depicted in Figure~\ref{fig:wsoft-outgoing-sum}. As we do this we cross a wall each time that the line of the soft particle can end on one of the D3 branes associated to the outgoing particles. The wall crossing formula for each of them will add a contribution similar to the right hand side of  \eqref{FirstCross} with $1 \to i$, and the soft particle ends on the $i^{th}$ outgoing particle. Because of \eqref{SoftCond}, the addition of the soft particle does not significantly change the slope of the other particles, and we do not have to worry about further possible wall crossing phenomena. 
In this way, we find
\be
\widetilde {\cal I}_n|_{\rm fig.\, \ref{fig:wsoft-outgoing-sum}}(1,\ldots,n-1,s)  
 =\sum_{j=1}^{n_{\rm out}}\langle \Gamma_j, \Gamma_s\rangle \,{\cal I}_{n-1}(1,\ldots,j+s,\ldots,n-1).
\ee

\FloatBarrier

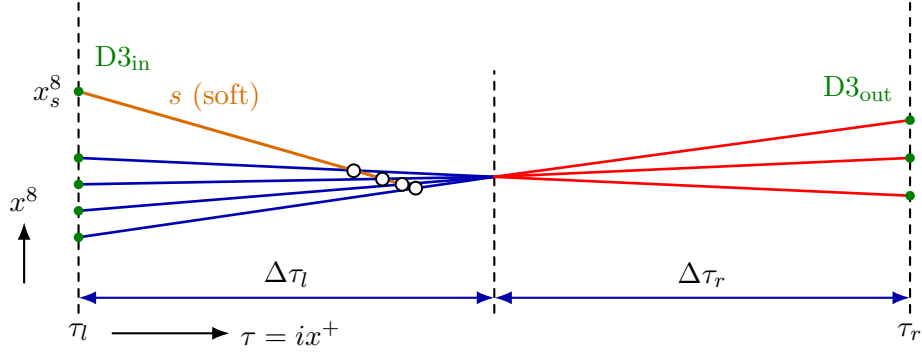
\begin{figure}[htbp]
  \centering
  \begin{tikzpicture}[
      x=1cm,
      y=1cm,
      line cap=round,
      line join=round,
      >=Latex
    ]
    \coordinate (hard) at (5.5,2.35);

    \draw[dashed, line width=0.8pt] (0,0.55) -- (0,4.65);
    \draw[dashed, line width=0.8pt] (5.5,0.55) -- (5.5,3.75);
    \draw[dashed, line width=0.8pt] (11,0.55) -- (11,4.65);
    \draw[blue!65!black, <->, line width=0.8pt]
      (0,0.75) -- node[above, text=black] {$\Delta\tau_l$} (5.5,0.75);
    \draw[blue!65!black, <->, line width=0.8pt]
      (5.5,0.75) -- node[above, text=black] {$\Delta\tau_r$} (11,0.75);
    \node[below] at (0,0.55) {$\tau_l$};
    \node[below] at (11,0.55) {$\tau_r$};

    \foreach \yy in {1.55,1.90,2.25,2.60} {
      \draw[blue!70!black, line width=1.0pt] (0,\yy) -- (hard);
      \fill[green!50!black] (0,\yy) circle[radius=1.8pt];
    }
    \foreach \yy in {2.10,2.60,3.10} {
      \draw[red, line width=1.0pt] (hard) -- (11,\yy);
      \fill[green!50!black] (11,\yy) circle[radius=1.8pt];
    }

    \draw[orange!85!black, line width=1.15pt] (0,3.48) -- (4.46,2.20);
    \fill[green!50!black] (0,3.48) circle[radius=1.8pt];
    \node[left] at (-0.08,3.48) {$x_s^8$};
    \node[above, orange!85!black] at (1.80,3.08) {$s\ \text{(soft)}$};

    \foreach \x/\y in {3.64/2.43,4.02/2.32,4.28/2.25,4.46/2.20} {
      \fill[white] (\x,\y) circle[radius=2.4pt];
      \draw[black, line width=0.75pt] (\x,\y) circle[radius=2.4pt];
    }

    \node[green!50!black, right] at (0.08,3.92) {$\mathrm{D3}_{\rm in}$};
    \node[green!50!black, left] at (10.92,3.52) {$\mathrm{D3}_{\rm out}$};
    \draw[->, line width=0.8pt] (-0.72,0.95) -- (-0.72,1.75)
      node[above] {$x^8$};
    \draw[->, line width=0.8pt] (0.45,0.28) -- (2.0,0.28)
      node[right] {$\tau=ix^+$};
  \end{tikzpicture}
  \caption{With the soft trajectory just below the central interaction point, raising $x_s^8$ crosses all $n_{\rm in}$ incoming legs. Each open circle denotes one possible trivalent junction and hence one term in the sum over incoming strings.}
  \label{fig:wsoft-incoming-sum}
\end{figure}

Similarly, starting from a very negative $x^8_s$ and moving the string up to just below the central point, one may conclude that for the string configuration in Figure~\ref{fig:wsoft-incoming-sum},
\be
 \widetilde{\cal I}_n|_{\rm fig.\,\ref{fig:wsoft-incoming-sum}}(1,\ldots,n-1,s) 
 =-\sum_{j=n_{\rm out}+1}^{n_{\rm out}+n_{\rm in}}\langle \Gamma_j , \Gamma_s \rangle \,
 {\cal I}_{n-1}(1,\ldots,j+s,\ldots,n-1).
\ee

\FloatBarrier

\begin{figure}[htbp]
  \centering
  \begin{tikzpicture}[
      x=1cm,
      y=1cm,
      line cap=round,
      line join=round,
      >=Latex
    ]
    \coordinate (hard) at (5.5,2.35);

    \draw[dashed, line width=0.8pt] (0,0.55) -- (0,4.85);
    \draw[dashed, line width=0.8pt] (5.5,0.55) -- (5.5,3.75);
    \draw[dashed, line width=0.8pt] (11,0.55) -- (11,4.85);
    \draw[blue!65!black, <->, line width=0.8pt]
      (0,0.75) -- node[above, text=black] {$\Delta\tau_l$} (5.5,0.75);
    \draw[blue!65!black, <->, line width=0.8pt]
      (5.5,0.75) -- node[above, text=black] {$\Delta\tau_r$} (11,0.75);
    \node[below] at (0,0.55) {$\tau_l$};
    \node[below] at (11,0.55) {$\tau_r$};

    \foreach \yy in {1.55,1.90,2.25,2.60} {
      \draw[blue!70!black, line width=1.0pt] (0,\yy) -- (hard);
      \fill[green!50!black] (0,\yy) circle[radius=1.8pt];
    }
    \foreach \yy in {2.10,2.60,3.10} {
      \draw[red, line width=1.0pt] (hard) -- (11,\yy);
      \fill[green!50!black] (11,\yy) circle[radius=1.8pt];
    }

    \draw[orange!85!black, line width=1.15pt] (0,3.93) -- (hard);
    \fill[green!50!black] (0,3.93) circle[radius=1.8pt];
    \node[left] at (-0.08,3.93) {$x_s^8$};
    \node[above, orange!85!black] at (2.05,3.43) {$s\ \text{(soft)}$};

    \node[green!50!black, right] at (0.08,4.38) {$\mathrm{D3}_{\rm in}$};
    \node[green!50!black, left] at (10.92,3.52) {$\mathrm{D3}_{\rm out}$};
    \draw[->, line width=0.8pt] (-0.72,0.95) -- (-0.72,1.75)
      node[above] {$x^8$};
    \draw[->, line width=0.8pt] (0.45,0.28) -- (2.0,0.28)
      node[right] {$\tau=ix^+$};
  \end{tikzpicture}
  \caption{At a special location in $x_s^8$,  between the configurations in Figure~\ref{fig:wsoft-outgoing-sum} and Figure~\ref{fig:wsoft-incoming-sum}, the soft string reaches the central interaction point.  There is no marginal-stability jump there because the would-be decay  products are not allowed. The resulting connected web gives the index ${\cal I}_n $ which computes the actual  $n$-point BFSS scattering amplitude.}
  \label{fig:wsoft-central}
\end{figure}
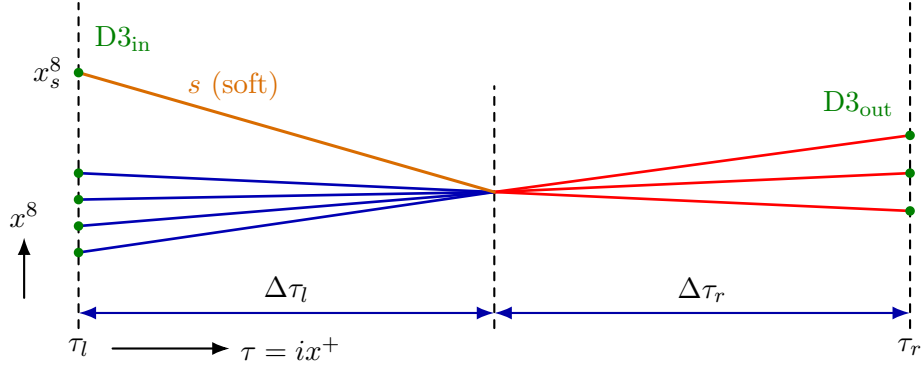

Directly sweeping  the soft string from 
 Figure~\ref{fig:wsoft-outgoing-sum} to Figure~\ref{fig:wsoft-incoming-sum}, one crosses the configuration in Figure \ref{fig:wsoft-central} representing the full $n$-point amplitude. However, there is no wall-crossing jump at this point. The would-be string web configuration would necessarily separate the soft string from the hard web, but neither object is allowed: the soft string has no allowed second endpoint, while the hard external momenta do not sum to zero. Hence, 
 the full $n$-point scattering amplitude illustrated in Figure~\ref{fig:wsoft-central} equals both that of Figure~\ref{fig:wsoft-outgoing-sum} and Figure~\ref{fig:wsoft-incoming-sum} which are therefore equal to one another
\begin{equation}\label{EqualityEqn}
 {\cal I}_n(1, \ldots, n-1,s) = \widetilde {\cal I}_n|_{\rm fig.\,\ref{fig:wsoft-outgoing-sum}}(1,\ldots,n-1,s) =\widetilde {\cal I}_n|_{\rm fig.\,\ref{fig:wsoft-incoming-sum}}(1,\ldots,n-1,s)
\end{equation} 
Alternatively, we can write it as the average of the two expressions as  
\be \label{eq:wsoft-ward}
 {\cal M}^{\rm BFSS}_n(1,\ldots,n-1,s)
=\frac{1}{2}
 \sum_{j=1}^{n-1}
  {\lvert}[sj] {\rvert}\,
 {\cal M}^{\rm BFSS}_{n-1}(1,\ldots,j+s,\ldots,n-1).
\ee

\FloatBarrier

This  is precisely the momentum-space Ward identity  of the $w_{1+\infty}$ symmetry of single-minus amplitudes.\footnote{  $w_{1+\infty}$ is the global subgroup of the loop group ${\cal L}w_{1+\infty}$ which is a symmetry of tree level gravity. Only this subgroup is relevant because  the dependence on the holomorphic spinors is trivial within a given chamber for single-minus amplitudes. It does not, however, tell us how to cross the walls from one chamber into another.  It is interesting to ask whether the full  loop group ${\cal L}w_{1+\infty}$ can do this job.}

Additionally, note that the second equality in \eqref{EqualityEqn} can also be written as 
\be \label{eq:soft-constraint}
0= \sum_{j=1}^{n-1}[sj]\,
  {\cal M}^{\rm BFSS}_{n-1}(1,\ldots,j+s,\ldots,n-1)\,.
 \ee
This cancellation is not ordinary momentum conservation, since the shifted lower-point amplitude depends on $j$. Rather it follows from the absence of a wall at the central interaction point or equivalently from the equality of the two wall-crossing paths.\footnote{
 This relation is implicit, but not explicitly displayed in the gravitational analysis of~\cite{GLSSW}. In Appendix~C of that paper, equations~(79) and~(86) show that the two sign sectors of the absolute-value Ward sum each equal $ {\cal M}_n$, while equation~(87) adds them to obtain $2{\cal M}_n$. Taking the difference instead, and using the fixed bracket signs in the ordered chamber, gives the above identity.} 

In this derivation, we put a restriction on the slope of the soft particle. Now we mention that the derivation also works without that restriction.  
We also start from a situation with $x_s^8 $ very large and then bring it down until the soft particle intersects the interaction point. As it does this, it will intersect all the lines with $[sj]>0$ (and only these lines); some of these are incoming lines and some are outgoing lines.  Similarly, starting from a very negative $x^8_s$ and increasing it until the soft particle intersects the interaction point will also intersect all the lines with $[sj]<0$ in the process. The final conclusion is similar, we get \eqref{eq:wsoft-ward}, as well as \eqref{eq:soft-constraint}.

Recall that we assumed \eqref{SoftCond} in order to prove \eqref{eq:wsoft-ward}. We assumed this so that all the terms in the right hand side of \eqref{eq:wsoft-ward} are in the same kinematic chamber. So we have shown that the equation is valid exactly (and not just to leading order in the expansion) for small enough $n_s$ and $N_s$. In other words, it is valid in an open neighborhood of $n_s=N_s=0$. 

\section{Scattering off a plane wave}
\label{sec:pp-wave-check}

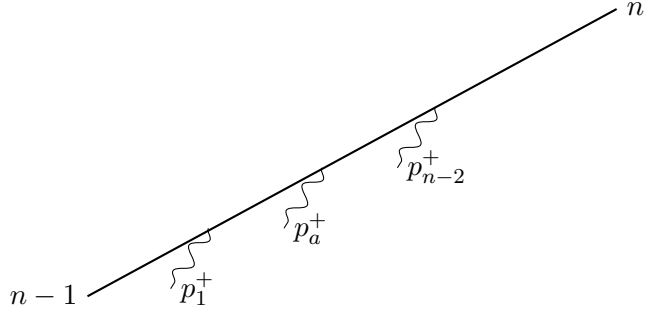
\begin{figure}[htbp]
\centering
\begin{tikzpicture}[scale=1.0, decoration={snake,amplitude=1.0mm,segment length=4mm}]
  \draw[thick] (0,0)--(7,3.8);
  \node[left] at (0,0) {$n-1$};
  \node[right] at (7,3.8) {$n$};
  \draw[decorate] (1.6,0.9)--(1.1,0.1);
  \draw[decorate] (3.1,1.7)--(2.6,0.9);
  \draw[decorate] (4.6,2.5)--(4.1,1.7);
  \node[right] at (1.1,0.1) {$p_1^+$};
  \node[right] at (2.6,0.9) {$p_a^+$};
  \node[right] at (4.1,1.7) {$p_{n-2}^+$};
\end{tikzpicture}
\caption{Sketch of the scattering configuration. Particles $n-1$ and $n$ are the only ones with nonzero $p_-$. We can view this as the propagation of a particle with nonzero $p_-$ through a complex plane wave created by the $n-2$ particles with zero $p_-$.}
\label{fig:app-ppwave-worldline}
\end{figure}

In this section, we consider an especially  simple and illuminating  single-minus scattering configuration. It leads to a simple picture in both gravity and the matrix model. It is a scattering process that can be viewed as a particle with nonzero $p_-$ propagating through a complex plane wave; see figure \ref{fig:app-ppwave-worldline}.
The second subsection gives the matrix-model version of this computation.

\subsection{A very simple single-minus amplitude: scattering off a plane wave}
\label{pwScatt}

We write the Minkowski metric as 
\begin{align}
    ds^2 = - 2\,\mathrm{d}x^+\,\mathrm{d}x^- + 2\,\mathrm{d}z\,\mathrm{d}\bar{z}.
\end{align} 
We take the first $n-2$ outgoing legs to be particles for which only the $p_z$ component of the momentum is nonvanishing.\footnote{These modes differ from the BFSS soft modes studied in~\cite{MSTW}, which correspond to RR fields in the IIA picture.}
The last two particles will have a nonzero $p_-$ component of the momentum,  which we will write in terms of $p_{\rm h}>0$. 
In summary, the momenta of all the particles are 
\begin{align}
  \begin{aligned}
    a=1,\ldots,n-2:\qquad
    &(p_{a-},p_{az},p_{a\bar{z}},p_{a+})
    \longrightarrow(0,q_a,0,0),\\
    n-1:\qquad
    &(p_{n-1,-},p_{n-1,z},p_{n-1,\bar{z}},p_{n-1,+})
    \longrightarrow(-p_{\rm h},0,0,0),\\
    n:\qquad
    &
    (p_{n-},p_{nz},p_{n\bar{z}},p_{n+})
    \longrightarrow(p_{\rm h},-Q,0,0),\qquad Q \equiv \sum_{a=1}^{n-2} q_a~.
  \end{aligned}
  \label{eq:blue-pp-limit}
\end{align}
Using spinor-helicity variables, we write the momentum and the positive- and negative-helicity polarization vectors as
\begin{align}\label{ConvMom}
p_{\alpha \dot \beta } = \left( \begin{array}{cc} -p_- & p_{  z} \\ -p_{\bar{z}} & p_+ \end{array}\right) = \lambda_{\alpha } \tilde{\lambda}_{\dot \beta } ~,~~~~~~ \epsilon^+_{\alpha \dot \beta } = \frac{\eta_\alpha\tilde{\lambda}_{\dot\beta}}{\langle\eta,\lambda\rangle}~,~~~~~~\epsilon^-_{\alpha \dot \beta } = \frac{\lambda_\alpha\tilde{\eta}_{\dot\beta}}{[\tilde{\eta},\tilde{\lambda}]}~.
\end{align}
\begin{align}
  \lambda_i=
  \begin{pmatrix}1\\ 0\end{pmatrix},
  \qquad
  \tilde{\lambda}_a=
  \begin{pmatrix}0\\ q_a\end{pmatrix},
  \qquad
  \tilde{\lambda}_{n-1}=
  \begin{pmatrix}p_{\rm h}\\ 0\end{pmatrix},
  \qquad
  \tilde{\lambda}_n=
  \begin{pmatrix}-p_{\rm h}\\ -Q\end{pmatrix},
  \quad a=1,\ldots,n-2 .
  \label{eq:blue-pp-spinors}
\end{align}
Notice that the momenta \eqref{eq:blue-pp-limit} are very special, even among the special momenta that lie on the support of the single-minus amplitude. In fact, they lie at the boundary of the chamber $\mathcal{R}_{n,n-1}$, approached from its
interior.
Among the brackets that enter 
\eqref{eq:IntroProduct}, the only surviving ones are
\begin{align}
  \begin{aligned}
    \relax[n-1,a]
 =-p_{\rm h}q_a ,
    \qquad a=1,\ldots,n-2 .
 \end{aligned}
  \label{eq:blue-pp-brackets}
\end{align}
Thus, every factor in \eqref{eq:IntroProduct} retains only
$[n-1,a]$, and the decay-chamber formula simplifies to
\begin{align}
  \left.
  \mathcal M^{\rm grav}_{n,+\cdots+-}
  \right|_{\rm pp}
  =
  \prod_{a=1}^{n-2}[n-1,a]
  =
  (-p_{\rm h})^{n-2}\prod_{a=1}^{n-2}q_a .
  \label{eq:blue-pp-product}
\end{align}

We now show that \eqref{eq:blue-pp-product} can be reproduced
from the worldline action of a graviton with nonzero $p_-$ propagating in the background of a plane wave.  
The plane wave has the form
\begin{align}
      ds^2=- 2\,\mathrm{d}x^+\,\mathrm{d}x^- + 2\,\mathrm{d}z\,\mathrm{d}\bar{z}+h_{++}(x^+,z)\left(\mathrm{d}x^+\right)^2,\qquad
      h_{++}=2\kappa_{11}\sum_{a=1}^{n-2}\epsilon_{++}^{(a)}e^{iq_az+ip_{a+}x^+}.
  \label{eq:blue-pp-metric}
\end{align}
Notice that the function $h_{++}$ is given by the superposition of the first $n-2$ gravitons in \eqref{eq:blue-pp-spinors} with positive helicity, obtained from \eqref{ConvMom} by choosing $\eta\propto\begin{pmatrix}0\\1\end{pmatrix}$. We have also temporarily included nonzero $p_{a+}$; these will be set to zero when we compute the amplitude. Note that \eqref{eq:blue-pp-metric}
 is Ricci-flat and supersymmetric:
$R_{++}=-\partial_z\partial_{\bar{z}}h_{++}=0$, although the Riemann tensor is nonzero,
$R_{+z+z}=-\frac12\partial_z^2h_{++}\neq0$.

We can think of the propagation of a graviton through this plane wave in the worldline formalism. We can further choose a light-cone gauge in which the particle action takes the form
\begin{align}\label{MPAction}
I = \frac{|p_-|}{2}  \int dx^+  \left[ 2 |\dot z|^2 + h_{++}(x^+,z,\bar{z}) 
\right]~.
\end{align} 
We can now compute the amplitude by expanding perturbatively to first order in each of the gravitons in $h_{++}$. The result is that we need to consider $n-2$ separate integrals along the worldline 
\begin{align}\label{Amplit}
 \mathcal{A} =\langle F| \prod_{a=1}^{n-2}\kappa_{11} |p_-| \int dx^+_a \epsilon_{++}^a e^{ i p_{a+} x_a^+ } e^{ i q_a z(x^+_a) } | I \rangle 
 \propto  p_{\rm h}^{n-2} \left[\prod_{a=1}^{n-2} q_a^2 \delta( p_{a + } ) \right]2p_{\rm h}\delta( \sum_{i=1 }^n p_{z i} )
 \end{align} 
where we used that each positive-helicity polarization vector in \eqref{ConvMom} is proportional to $q_a$, using \eqref{eq:blue-pp-spinors}. The last factor $(2p_{\rm h})$ in \eqref{Amplit} comes from the overlap of $|I\rangle$ and $|F\rangle$, which are the initial and final states of the particle with nonzero $p_-=\pm p_{\rm h}$.
 We also used that the correlators of $z(x^+_a)$ depend only on the zero mode but not on the time $x^+$, since there are no contractions among the $z$'s (only between $z$ and $\bar{z}$). The integral over the $z$ zero mode produces the last delta function in the expression. There are other obvious momentum-conserving delta functions that we are not indicating yet. 
We see that this produces the extra delta functions present in the single-minus amplitude. These delta functions can also be viewed as enforcing that all $\lambda_a$ are equal. In fact, when $p_{a+}$ is nonzero, the spinor-helicity variable becomes
 \begin{align}
 \lambda_a =  \begin{pmatrix}1\\  p_{a+}/q_a \end{pmatrix} =\begin{pmatrix}1\\  \lambda_{2a} \end{pmatrix}.
 \end{align} 
 Therefore $|q_a| \delta(p_{a+}) =  \delta( \lambda_{2a} ) $ and restoring the prefactor and extra delta functions we find 
 \begin{align}\label{AmplitFin}
 \mathcal{A} = 2\kappa_{11}^{n-2}p_{\rm h}^{n-2} (2\pi)^3 \left[\prod_{a=1}^{n-2} 2\pi |q_a|\delta( \langle a n \rangle ) \right]\delta( \sum_{i=1 }^n p_{z i} ) \delta ( \sum_{i=1}^n p_{- i })    \delta ( \langle n-1,n\rangle )
 \end{align} 
where the factor in brackets comes from the $(n-2)$ integrals $\int dx_a^+$ discussed in \eqref{Amplit}. The delta function for $\sum_i p_{zi}$ comes from the $z$ zero-mode integral and includes the contributions of the initial and final worldline states. The delta function for $\sum_i p_{-i}$ receives contributions only from particles $n-1$ and $n$. The final delta function in \eqref{AmplitFin} follows from momentum conservation in the $\bar{z}$ direction after a small shift in the second component of $\lambda_{n-1} \propto p_{\bar{z},n-1}$. 
 More precisely, on the support of the first $n-2$ delta functions $p_{a+}=0$ one shows that 
\begin{equation}
    p_{\rm h} \delta( \sum_{i=1 }^n p_{\bar{z} i} ) = \delta(\langle n-1,n \rangle )\,.
\end{equation}
Thus \eqref{AmplitFin} reproduces the single-minus graviton component of the superamplitude \eqref{eq:IntroFullAmplitude} with the prefactor \eqref{eq:blue-pp-product}. The two delta functions involving sums are precisely those obtained from the sum over the $\tilde{\lambda}_{i\dot\beta}$ variables in the last factor of \eqref{eq:IntroFullAmplitude}. Thus, for this particular kinematics, the origin of both the delta functions and their prefactor is clear.
 
\subsection{The matrix-model computation}

We will now consider the corresponding matrix-model computation. This is particularly simple because the first $n-2$ momenta have zero $p_-$. Thus, the problem reduces to a matrix-quantum-mechanics problem where we add the function $h_{++}$ as an external potential after promoting the coordinates to matrices. In other words, we take $p_{\rm h}=N/R_-$ and consider the action
\begin{align}
I = \frac{1}{2R_-} \int dx^+ Tr[ 2 \dot Z \dot {\bar{z} } + h_{++}(x^+, Z ) ]
\end{align} 
where now $Z$ is an $N\times N$ matrix; see~\cite{TaylorVanRaamsdonk}. This is completely analogous to the action \eqref{MPAction}. Again, we expand the function $h_{++}$ to first order for each graviton. The main feature is that only the $\mathsf{U}(1)$ center-of-mass degree of freedom contributes to the computation, which becomes identical to the one in \eqref{Amplit}. In other words, we write
\begin{align} 
Z = z {\bf 1 }_N  + \hat Z 
\end{align} 
where $\hat Z$ is an $\mathsf{SU}(N)$ matrix and ${\bf 1 }_N $ is the identity matrix. Then the action for $z$ and the corresponding computation of the amplitude is the same as what we did for the gravity case in section~\ref{pwScatt}. In order to reach this conclusion, we
used that all nonconstant terms in the expansion of the $\mathsf{SU}(N)$ holomorphic correlators vanish in the ground state, so that
\begin{align}
\frac{1}{R_-^{n-2}} \langle 0 |\prod_{a=1}^{n-2} \operatorname{Tr}[e^{ i q_a \hat Z(x^+_a)}] |0\rangle =
\frac{N^{n-2}}{R_-^{n-2}}  = p_{\rm h}^{n-2}
\end{align}
where $\hat Z$ is the $\mathsf{SU}(N)$ part of the matrix $Z$.  This simply follows from the fact that $\hat Z$ carries a nontrivial $\mathsf{U}(1) $ charge under rotations in the $Z,\bar{z}$ plane. This argument is subtle because the ground state has power law tails, while $e^{ i q Z} $ grows exponentially along the $X^8$ direction. So we are assuming that we have a rotation invariant way to regularize the integrals (or define them by a suitable analytic continuation).
 
Scattering in more general self-dual
radiative backgrounds was studied in~\cite{AdamoMasonSharma, Guevara:2024edh}. Such backgrounds allow non-collinear chiral
data and might probe the full single-minus structure.

\section*{Acknowledgments}

We are grateful to David Skinner for useful conversations.
A.G. thanks Aidan Herderschee for discussions about the BFSS model,  and acknowledges the Roger Dashen Membership at the IAS.
A.G. and J.M. are supported in part by the U.S. Department of Energy grant DE-SC0009988. 
J.M. is part of the Leinweber Forum at IAS. 
A.L. is supported in part by NSF grant AST-2307888, NSF CAREER award PHY-2340457, and the Simons Foundation grant SFI-MPS-BH-00012593-09.
A.S. is supported in part by the U.S. Department of Energy grant DE-SC0007870.

\textbf{Statement on use of AI:} The ideas, arguments and writing on this paper came from the human authors.
An internal OpenAI model provided useful assistance with aspects of the analysis, checking numerical factors, and copy editing.

\appendix

\section{Three-point normalization check}
\label{app:three-point-normalization}

In this Appendix, we briefly discuss the $n=3$ amplitude. The purpose is two-fold: On the one hand this illustrates the conventions used throughout the text, and on the other, we show how to massage the standard spinor-helicity expression into the frame \eqref{eq:IntroFullAmplitude}. 

The Feynman-normalized three-point superamplitude is~\cite{HM3}
\begin{equation} \label{eq:app-three-superamplitude}
{\cal A}_3
=2\kappa_{11}(2\pi)^4\delta^4(P)
\frac{\delta^8\!\left(
 \widetilde\eta_1[23]+\widetilde\eta_2[31]
 +\widetilde\eta_3[12]\right)}
{[12]^2[23]^2[31]^2},
\qquad P=\sum_{i=1}^3p_i .
\end{equation}

In the fixed gauge of \eqref{eq:frcho}, the two rows of the
total momentum matrix are
\begin{equation} \label{eq:app-three-momentum-matrix}
\begin{aligned}
P_{1\dot\alpha}&=\sum_i\widetilde\lambda^i_{\dot\alpha},\\
P_{2\dot\alpha}&=z_3\sum_i\widetilde\lambda^i_{\dot\alpha}
 +\langle13\rangle\widetilde\lambda^1_{\dot\alpha}
 +\langle23\rangle\widetilde\lambda^2_{\dot\alpha}.
\end{aligned}
\ee
After the first row is set to zero, the Jacobian with respect to
\((\langle13\rangle,\langle23\rangle)\) is
\begin{equation}\label{eq:app-three-jacobian}
\left|\det
\begin{pmatrix}
\widetilde\lambda^1_{\dot1}&\widetilde\lambda^2_{\dot1}\\
\widetilde\lambda^1_{\dot2}&\widetilde\lambda^2_{\dot2}
\end{pmatrix}\right|
=|[12]|,
\end{equation}
and therefore
\begin{equation} \label{eq:app-three-support}
(2\pi)^4\delta^4(P)
=\frac{
 [2\pi\delta(\langle13\rangle)]
 [2\pi\delta(\langle23\rangle)]}
 {|[12]|}
 (2\pi)^2\delta^2\!\left(\sum_i\widetilde\lambda_i\right).
\end{equation}

Dotted momentum conservation implies
\begin{equation} \label{eq:app-three-brackets}
\widetilde\lambda_3=-\widetilde\lambda_1-\widetilde\lambda_2,
\qquad
[23]=[31]=[12]=:s .
\ee
The kinematic factor in \eqref{eq:app-three-superamplitude}, including the
Jacobian in \eqref{eq:app-three-support}, consequently reduces to
\begin{equation} \label{eq:app-three-reduction}
\begin{aligned}
\frac{\delta^8\!\left(s\sum_i\widetilde\eta_i\right)}{s^6|s|}
&=\frac{s^8}{s^6|s|}
  \delta^8\!\left(\sum_i\widetilde\eta_i\right)\\
&=|s|\,\delta^8\!\left(\sum_i\widetilde\eta_i\right).
\end{aligned}
\end{equation}
For \(n=3\), \eqref{eq:IntroProduct} gives
\({\cal M}^{\rm grav}_3=|[12]|=|s|\).  Combining
\eqref{eq:app-three-superamplitude},
\eqref{eq:app-three-support}, and
\eqref{eq:app-three-reduction} gives
\begin{equation} \label{eq:app-three-result}
\begin{aligned}
{\cal A}_3={}&2\kappa_{11}{\cal M}^{\rm grav}_3
[2\pi\delta(\langle13\rangle)]
[2\pi\delta(\langle23\rangle)]\\
&\times(2\pi)^2\delta^2\!\left(\sum_i\widetilde\lambda_i\right)
\delta^8\!\left(\sum_i\widetilde\eta_i\right).
\end{aligned}
\ee
This is exactly the \(n=3\) specialization of
\eqref{eq:IntroFullAmplitude}, \eqref{Intro34pt}.

\end{document}